\documentstyle {article}
\title {Grassmannian and string theory.}
\author {Albert Schwarz\\ Department of Mathematics,
University of California,\\Davis, CA 95616\\
SCHWARZ@MATH.UCDAVIS.EDU}
\begin {document}
\maketitle
\smallskip
 \begin{abstract}

  Infinite-dimensional Grassmannian manifold contains moduli 
spaces of Riemann surfaces of all genera. This well known fact 
leads to a conjecture that non-perturbative string theory can 
be formulated in terms of  Grassmannian. We present new facts 
supporting this hypothesis. In particular, it is shown that 
Grassmannians can be considered as generalized moduli spaces; 
this statement permits us to define corresponding "string 
amplitudes" (at least formally). One can conjecture, that it 
is possible to explain the relation between non-perturbative 
and perturbative string theory  by means of localization theorems 
for equivariant cohomology; this conjecture is based on the 
characterization of moduli spaces, relevant to string theory, 
as sets  consisting of points with large stabilizers in certain 
groups acting on  Grassmannian. We describe an involution on the  
Grassmannian that could be related to $S$-duality in string theory.

\end{abstract} 

{\bf 0. Introduction.}
\vskip .1in
  It is clear now that all versions of string theory 
are closely related.One should expect that all of them 
can be obtained from a unifying theory, where  strings 
are not considered as fundamental objects (strings should 
be on equal footing with membranes). The present paper 
arose from attempts to understand the structure of the 
unifying theory. If we believe that the analysis of this 
hypothetical theory is related to calculation of integrals 
over infinite-dimensional supermanifolds then we can 
suggest the following picture. The integrand can have 
various odd symmetries. Under certain conditions the 
existence of odd symmetry leads to localization; in other 
words, one can replace the integration over the whole 
supermanifold with an integration over some part of it. 
(See [11],[13],[14],[15] for localization theorems in the 
framework of supergeometry and of equivariant cohomology.)
Different odd symmetries lead to different localizations; 
and therefore to theories that could look completely unrelated.

  A very natural candidate for infinite-dimensional 
supermanifold arising in the universal theory is 
infinite-dimensional (super)Grassmannian. I hope that 
appropriate integrals could be localized to the Krichever 
locus; such a localization would lead immediately to 
relation to the string theory. I have only tentative 
results in this direction. However, I was able to prove 
that Grassmannians can be considered as generalized moduli 
spaces, containing many other moduli spaces, and to 
describe a generalization of string theory that is defined 
in terms of Grassmannian.

  Bosonic string theory is closely related to two-dimensional 
conformal field theory, other versions of string theory  are 
related to corresponding generalizations of conformal field 
theory. Let us consider the moduli space $P_{\chi ,n}$ of (possibly,disconnected) compact conformal two-dimensional 
manifolds of Euler characterictic $\chi$ having a boundary 
consisting of $n$ components (i.e. closed two-dimensional 
manifolds of Euler characteristic $\chi -n$ with $n$ holes). 
We assume that boundaries of holes are parametrized and the 
holes are ordered. One can define natural maps $\nu_{n,n^{\prime}}: 
P_{\chi ,n}\times P_{\chi^{\prime},n^{\prime}}\rightarrow 
P_{\chi +\chi ^{\prime},n+n^{\prime}}$ and $\sigma ^{(n)}: 
P_{\chi ,n}\rightarrow P_{\chi ,n-2}$ (the first map corresponds 
to disjoint union of manifolds, the second one is constructed 
by means of pasting together of the boundary of $(n-1)$-th hole 
and the boundary of $n$-th hole). The symmetric group $S_n$ acts 
naturally on $P_{\chi ,n}$ (reordering the holes).

  Sometimes it is useful to take into account that pasting 
together boundaries of two holes we can rotate one of the 
boundaries. Then we obtain a map $P_{\chi ,n}\times S^1 
\rightarrow P_{\chi ,n-2}$.

  The situation when we have some spaces $P_n$ with action 
of symmetric group $S_n$ and maps $\nu _{n,n^{\prime}}: 
P_n\times P_{n^{\prime}}\rightarrow P_{n+n^{\prime}},\  
\sigma ^{(n)}:P_n\rightarrow P_{n-2}$ appears quite often. 
If these data satisfy some compatibility conditions we will 
say that the spaces $P_n$ constitute a MO (a modular operad). 
The spaces $P_n=\cup _{\chi}P_{\chi,n}$ form a MO; we 
reflect the fact $P_n$ has such a decomposition, saying 
that we have a graded MO. If we include in our data also 
a map $P_n\times S^1\rightarrow P_{n-2}$, we will talk 
about an EMO (equivariant MO). The spaces $P_{\chi ,n}$ 
constitute a graded EMO.

  If $A$ is a linear space with inner product we can define 
a MO considering tensor powers $A^{\otimes n}$ and natural 
maps $A^{\otimes n}\times A^{\otimes n^{\prime}}\rightarrow 
A^{\otimes (n+n^{\prime})}$ and $A^{\otimes n}\rightarrow 
A^{\otimes (n-2)}$. An algebra over the MO $P_n$ is defined 
as a collection of maps $\alpha _n:P_n\rightarrow A^{\otimes 
n}$, that are compatible with the structures of MO in $P_n$ 
and $A^{\otimes n}$. Conformal field theory can be defined 
as an algebra over the MO $P_n=\cup P_{\chi ,n}$, described 
above. (More precisely, in this case the maps $\alpha _n$ 
are defined only up to a constant factor.) A superconformal 
field theory can be defined in similar way. Moduli spaces of 
conformal  manifolds (of complex curves) should be replaced 
in this case with moduli spaces of superconformal manifolds 
(super Riemann surfaces).

  We will be able to construct a lot of MOs starting with 
an infinite-dimensional Grassmannian.

  We will say that a semi-infinite structure in Hilbert space 
$H$ is specified by a decomposition $H=H_+\oplus H_-$ and a 
unitary involution $K$ interchanging $H_+$ and $H_-$. The 
Segal-Wilson modification of Sato Grassmannian $Gr(H)$ can be 
defined as an infinite-dimensional manifold consisting of 
such subspaces $V\subset H$ that $V$ is close to $H_+$ in 
some sense (one can require for example that the projection 
$V\rightarrow H_+$ is a Fredholm operator and the projection 
$V\rightarrow H_-$ is a Hilbert-Schmidt operator).

  We will prove that the sequence of spaces $Gr_n=Gr(H^n)$ can 
be considered as MO. Moreover, we will give conditions when 
$Gr_n$ constitute an EMO. Using well known fermionic construction 
of Grassmannian we will obtain an algebra over this MO.

  Let us consider now an infinite-dimensional algebra ${\cal G}$ 
acting on $H$; then ${\cal G}^n$ acts on $H^n$ and, under certain 
conditions, on $Gr(H^n)$. We define ${\cal G}$-locus $P_n({\cal G})$ 
as a subspace of $Gr(H^n)$ consisting of subspaces $V\subset H^n$ 
having "large stabilizers" in ${\cal G}$. (More precisely, we 
introduce a semi-infinite structure in ${\cal G}$ and require 
that the stabilizer $Stab_V$ contains a space $W\in Gr({\cal G}^n)$.) 
We prove that the spaces $P_n({\cal G})$ also can be used to 
construct MO. For appropriate choice of ${\cal G}$ we can consider 
conformal field theories, WZNW models etc as algebras over 
$P_n({\cal G})$. Replacing Grassmannian with super Grassmannian 
we can obtain also superconformal theories.

  It is well known that a conformal field theory can be 
considered as a background for a string theory and that 
one can construct corresponding string amplitudes as 
integrals over moduli spaces (at least in the case of 
critical central charge.) Generalizing these constructions 
we can define (at least formally) "string amplitudes"  
corresponding  to an algebra over an EMO. It seems that 
the "string amplitudes" corresponding to $Gr(H^n)$ 
could be universal in some sense. We describe an involution 
on $Gr(H^n)$ that could be related to $S$-duality in string 
theory. (The space $Gr(H^n)$ can be represented as a union 
of connected  components $Gr^{(k)}(H^n)$, where $k$ stands for 
the index 
of Fredholm operator $V\rightarrow H_+$, corresponding to 
$V\in Gr(H^n)$. It is natural to conjecture that in 
perturbation theory the contribution of $Gr^{(k)}(H^n)$ is 
proportional to $g^k$, where $g$ stands for the coupling 
constant. The involution we constructed transforms $Gr^{(k)}(H^n)$ 
into $Gr^{(-k)}(H^n)$.)

{\bf 1. Grassmannian}
\vskip .1in
   Let us consider a (complex) Hilbert space $H$ represented 
as a direct sum of subspaces $H_+$ and $H_-$. We will assume 
that there exists a unitary involution $K$ on $H$ transforming 
$H_+$ into $H_-$ and  $H_-$ into $H_+$. We will say that 
subspaces  $H_+, H_-$ and the involution $K$ specify a 
semi-infinite structure on $H$. As an example, we can take 
$H=L^2(S^1)$ where $S^1$ is the unit circle in $\bf C$ with 
standard measure. Then we can define $H_-$ as a subspace 
spanned by $z^n=e^{in\varphi },\  n\geq 0$ and $H_+$ as a 
subspace spanned by $z^n=e^{in\varphi},\  n<0$. The 
involution $K$ can be chosen as a map transforming a 
function $f(z)$ into a function ${1\over z}f({1\over z})$. 
It is easy to check that every semi-infinite structure is 
isomorphic to the standard structure described above. (Notice, 
that we include separability and infinite-dimensionality in 
the definition of Hilbert space. Subspaces are by definition 
closed linear submanifolds.) Using a semi-infinite structure 
in Hilbert space $H$ we can define a Grassmannian $Gr(H)$ as 
a set of such subspaces $V\subset H$ that the natural 
projection $\pi _+$ of $V$ into $H_+$ is a Fredholm operator 
and the natural projection $\pi _-$ of $V$ into $H_-$ is a 
compact operator. The  Grassmannian $Gr(H)$ can be represented 
as a union of connected components $Gr^{(k)}(H)$ were $k$ 
stands for the index of the Fredholm operator $\pi _+: 
V\rightarrow H_+$.

  Let us consider a map $\alpha :H_+\rightarrow H=H_+\oplus H_-$ 
transforming $h\in H_+$ into $\alpha (h)=(A+a)h+Bh$, where $A$ is 
an invertible operator acting on $H_+$ and $a:H_+\rightarrow H_+,\
B:H_+\rightarrow H_-$ are compact operators. It is easy to check 
that the space $\alpha (H_+)\subset H$ belongs to  $Gr^{(0)}(H)$ 
and that every $V\in Gr^{(0)}(H)$ can be obtained by means of this construction. This gives us an alternative description of 
$Gr^{(0)}(H)$. One can give a similar description of $Gr^{(k)}(H)$: 
if $F$ is a Fredholm operator of index $k$ acting on $H_+$ and 
$B$ is a compact operator acting from $H_+$ into $H_-$, then the 
image of the operator $\alpha :H_+\rightarrow H$, where $\alpha (h)
=Fh+Bh$, belongs to $Gr^{(k)}(H)$ and every element of $Gr^{(k)}(H)$ 
can be obtained this way.

  If $H$ has the form $H_+\oplus H_-$  a linear operator $A$ acting 
in $H$ can be represented in the form

$$\tilde {x}_+=A_{++}x_++A_{+-}x_-$$

$$\tilde {x}_-=A_{-+}x_++A_{--}x_-$$

where $x_+,\tilde {x}_+\in H_+,\  x_-,\tilde {x}_-\in H_-$. We will 
say that $A\in GL(H)$ if $A$ is an invertible operator, the operators
$A_{++}:H_+\rightarrow H_+$ and  $A_{--}:H_-\rightarrow H_-$ are
Fredholm, the operators $A_{+-}:H_-\rightarrow H_+,\
A_{-+}:H_+\rightarrow H_-$ are compact. The Lie algebra $gl(H)$ of
$GL(H)$ is defined as the set of operators $A$ obeying $\exp (At)\in
GL(H)$. It is easy to check that $GL(H)$ acts on $Gr(H)$ and that 
this action is transitive on every component $Gr^{(k)}(H)$.

  Notice, that a semi-infinite structure on $H$ induces naturally 
a semi-infinite structure on $H^n$ (on the direct sum of $n$ copies 
of $H$). Namely, $H^n=H\oplus ...\oplus H=(H_+\oplus ...\oplus H_+)
\oplus (H_-\oplus ...\oplus H_-)$; the involution $K:H^n\rightarrow 
H^n$ is a direct sum of $n$ copies of $K:H\rightarrow H$. Hence, we 
can speak about $Gr(H^n)$ and about $Gr^{(k)}(H^n)$.

  One can modify the definition of Grassmannian and of $GL(H)$
replacing compact operators with operators belonging to the trace 
class or with  Hilbert-Schmidt operators. All considerations of 
this section remain valid after such a modification. The set of 
compact  operators ${\cal B}$ can be considered as an ideal in 
the algebra $L(H)$ of bounded  operators; we can replace ${\cal B}$ 
with any other ideal consisting of compact  operators and 
containing all finite-dimensional operators.
 
 One can define also a semi-infinite structure on a normed space 
$H$. Such a structure is specified by means of subspaces $H_+$ 
and $H_-$ and an operator $K$, interchanging $H_+$ and $H_-$. 
(It is not necessary to assume that $K$ is an involution.) We 
assume that $K$ is an invertible operator on $H$ and that $H$ 
is equivalent to the direct sum $H_+\oplus H_-$ (more precisely, 
there exists an invertible operator $\pi =(\pi _+,\pi _-): 
H\rightarrow H_+\oplus H_-$ such that $\pi_{\pm}h=h$ for 
$h\in H_{\pm}$). Again we can define the Grassmannian $Gr(H)$ 
and prove in this more general situation the results stated below.

  Let us fix a space $H$ equipped with semi-infinite structure. 
We will use sometimes the notations $Gr,\  Gr_n$ and $Gr_n^{(k)}$ 
instead of $Gr(H),\  Gr(H^n)$ and $Gr^{(k)}(H^n)$ correspondingly. 
Let us consider linear subspaces $V\subset H^m$ and  $V^{\prime} 
\subset H^n$. Then their direct sum $V\oplus V^{\prime}$ belongs 
to $Gr(H^{m+n})$. We obtain a map

$$\nu_{m,n} :Gr(H^m)\times Gr(H^n)\rightarrow Gr(H^{m+n}).$$

For every subspace $V\subset H^m$ we can construct a subspace
$\sigma(V)\subset H^{m-2}$ consisting of points $(f_1,...,f_{m-2}) 
\in H^{m-2}$, satisfying the condition that one can find $u\in H$ 
in such a way that $(f_1,...,f_{m-2},Ku,u)\in V$.

  {\bf Theorem 1.} If $V\in Gr(H^m)$, then $\sigma(V)\in Gr(H^{m-2})$.

  To prove this statement we represent $V$ as the image of a 
map $H_+^m\rightarrow H^m$ transforming a point $(h_1,...,h_m) 
\in H_+^m$ into a point $(f_1+g_1,...,f_m+g_m)\in H^m$, where 
$f_k=\sum A_{kj}h_j\in H_+,\ \  g_k=\sum B_{kj}h_j\in H_-$, the 
operators $A_{kk}:H_+\rightarrow H_+$ are Fredholm, the
 operators  $B_{kj}:H_+\rightarrow H_-$ and operators $A_{kj}:H_+\rightarrow 
H_+,\  k\neq j$, are compact. To describe the space $\sigma (V)$ 
we impose the condition $f_{m-1}+g_{m-1}=K(f_m+g_m)$ on the points 
of $V$. Taking into account that $f_k\in H_+,\  g_k\in H_-$ we 
can rewrite this condition in the form

\begin {equation}
\begin {array} {rcl}
\sum_j A_{m-1,j}h_j & = & K\sum _jB_{mj}h_j\\
\sum _jB_{m-1,j}h_j & = & K\sum _jA_{mj}h_j
\end {array}
\end {equation}

  Now we can apply the following statement.

 {\bf Lemma 1.} Let us consider an equation

 \begin {equation}
 Fx=Ly
 \end {equation}
where $x,y$ are elements of infinite-dimensional Banach spaces $E$ and 
$E^{\prime}$. Assume that the operator $F:E\rightarrow E$ is Fredholm and 
the operator $L:E^{\prime}\rightarrow E$ is compact. Then one can 
find a Fredholm operator $G:E^{\prime}\rightarrow E^{\prime}$ 
and a compact operator $M:E^{\prime}\rightarrow E$ in such a way 
that for every element $u\in E^{\prime}$ the elements

\begin {equation}
x=Mu,\ \ y=Gu
\end {equation}
obey (2). Moreover, one can find $G$ and $M$ in such a way that 
every solution to (2) can be represented in the form (3) and 
this representation is unique. (Then index $G=-$ index $F$.)

   We apply Lemma 1 to Equation (1) taking $x=(h_{m-1},h_m),\
y=(h_1,...,h_{m-2})$, $E=H_+\oplus H_+,\  E^{\prime}=H_+^{m-2}$. 
The representation of $\sigma (V)$ obtained this way makes the 
statement of Theorem 1 obvious.

  Let us consider an infinite-dimensional Lie algebra ${\cal G}$ 
and a homomorphism $\alpha :{\cal G}\rightarrow gl(H)$ where $H$ 
is provided with  semi-infinite structure $(H_+,H_-,K)$. We 
assume that there exists an involution $\kappa$ of the Lie 
algebra ${\cal G}$ obeying $\alpha (\kappa (\gamma))=K\alpha 
(\gamma)K^{-1}$ and that one can find a  semi-infinite structure 
on ${\cal G}$ specified by means of subspaces ${\cal G}_+$ and 
${\cal G}_-$ and involution $\kappa$.

 The Lie algebra ${\cal G}^m$ (direct sum of $m$ copies of ${\cal G}$)
acts  on $H^m$ (an element $(\gamma _1,...,\gamma_m)\in {\cal
G}^m$ transforms  $(f_1,...,f_n)\in H^n$ into $(\alpha(\gamma
_1)f_1,...,\alpha(\gamma _m)f_m)\in H^m$). It is easy to check that this
action generates an action of ${\cal G}^m$ on $Gr(H^m)$. Let us denote
by $Stab_V$ a subalgebra of ${\cal G}^n$ consisting of elements
transforming an element $V\in Gr(H^m)$ into itself (the stabilizer of
$V$). We will define a ${\cal G}$-locus $P_m({\cal G})\in Gr(H^m)$ as a
set of such points $V\in Gr(H^m)$ that there exists an element $W\in
Gr({\cal G}^m)$ obeying $Stab_V\supset W$. In other words $P_m({\cal G})$
consists of elements $V\in Gr(H^m)$ having "large stabilizers" in
${\cal G}^m$.

 {\bf Theorem 2.}  If $V\in P_m({\cal G})$, then $\sigma(V)\in
P_{m-2}({\cal G})$.

 To prove this theorem we should check that $\sigma (V)$ has a large
stabilizer. If $(\gamma _1,...,\gamma_m)\in  Stab_V$ then for  
$(f_1,...,f_m)\in V$ we have  $(\alpha(\gamma _1)f_1,...,\alpha(\gamma
_m)f_m)\in V$. If we know that $(\gamma _1,...,\gamma_m)\in  Stab_V,\
\gamma_{m-1}=\kappa (\gamma _m)$ we can check that $(\gamma
_1,...,\gamma _{m-2})\in Stab _{\sigma (V)}$. To verify this statement
we take a point  $(f_1,...,f_{m-2})\in \sigma (V)$ constructed by
means of $(f_1,...,f_m)\in V$ where $f_{m-1}=Kf_m$. Using the  relation

  \begin {equation}
 \alpha(\gamma_{m-1})f_{m-1}=\alpha(\kappa(\gamma_m))Kf_m=K\alpha(\gamma
 _m)f_m
 \end {equation}
we see that the point

$$(\alpha(\gamma _1)f_1,...,\alpha(\gamma _{m-2})f_{m-2})$$

belongs to $\sigma (V)$ (one can construct it using the point
$(\alpha(\gamma _1)f_1,...,\alpha(\gamma _m)f_m)\in V$). This 
means that  $Stab_{\sigma (V)}\supset \sigma (Stab _V)$. If 
$Stab_V\supset W\in Gr({\cal G}^m)$ we can conclude from 
Theorem 1 that $Stab _{\sigma (V)}\supset \sigma (W)\in 
Gr({\cal G}^{m-2})$ and therefore $\sigma (V)\in P_{m-2}({\cal G})$.

  The following statement is almost evident:

  If $V\in P_m({\cal G}),\  V^{\prime}\in P_n({\cal G})$, then 
$V\oplus V^{\prime}\in P_{m+n}({\cal G})$.

 We obtain a map $\nu_{m,n}:P_m({\cal G})\times P_n({\cal G}) 
\rightarrow P_{m+n}({\cal G})$.

 Let us consider a compact one-dimensional complex manifold 
(complex curve) $\Sigma$ and $s$ holomorphic maps of the 
standard disk $D=\{ z\in {\bf C}\mid |z|\leq 1\}$ into $\Sigma$. 
We assume that the images $D_1,...,D_s$ of these maps do not 
overlap. Let us fix a holomorphic line bundle $\xi$ over 
$\Sigma$ and trivializations of this bundle over the disks 
$D_1,...,D_s$. We denote by $W=W(\Sigma,\xi )$ the space of 
holomorphic sections of the bundle over $\Sigma \setminus 
(D_1\cup ...\cup D_s)$. Restricting a section $s$ of $\xi$ to 
the boundaries $\partial D_1,...,\partial D_s$ and using the 
trivializations of $\xi$ over the disks we obtain an element
$\rho (s)$ of $H^s$, where $H$ stands for the space $L^2(S^1),\  
S^1$ denotes the standard circle $|z|=1$. The image $\sigma (W)$ 
of the space $W$ is a linear subspace of $H^s$; one can check [1] 
that $\sigma (W)\in Gr(H)$. (We equip $H^s$ with the standard 
semi-infinite structure.) One can generalize this construction 
allowing $\Sigma$ to be a complete irreducible complex algebraic 
curve (possibly singular) and replacing the line bundle $\xi$ 
with rank $1$ torsion free coherent sheaf on $\Sigma$. (If 
$\Sigma$ is non-singular such a sheaf is a vector bundle.) Then 
again one can prove that  $\sigma(W)\in Gr(H^s)$. The points 
$\sigma (W)$ obtained by means of the  construction above 
constitute so called Krichever locus in $Gr(H^s)$.

 Let us consider now the set $\Gamma$ consisting of invertible 
twice differentiable functions on $S^1$. Every element $\gamma 
\in \Gamma$ generates an operator $\alpha (\gamma):H\rightarrow 
H$ transforming $f\in H$ into $\gamma f$; it is easy to check 
that $\alpha (\gamma)\in GL(H)$ [1]. Considering $\Gamma$ as a 
group with respect to multiplication we can say that $\alpha$ 
is a homomorphism of the group $\Gamma$ into $GL(H)$.

  Let us denote by ${\cal A}_{\Sigma}$ the space of invertible
holomorphic functions on $\Sigma \setminus (D_1\cup ...\cup D_s)$.
Restricting these functions to the boundaries   $\partial
D_1,...,\partial D_s$  we obtain an embedding of ${\cal A}_{\Sigma}$
into $\Gamma ^s$; the image of this embedding will be denoted by ${\cal
A}_{\Sigma}^{\Gamma}$. It is easy to see that ${\cal A}
_{\Sigma}^{\Gamma}V\subset V$ for every $V\in Gr(H^s)$ obtained by
means of Krichever construction starting with $\Sigma ,D_1,...,D_s$
(for any choice of the bundle or sheaf $\xi$ and trivializations of
it). The Lie algebra Lie $\Gamma$ of the group $\Gamma$ consists of all
twice differentiable functions on the circle; the Lie algebra $({\rm
Lie}\ \Gamma)^s={\rm Lie}\ \Gamma^s $ can be represented by means of
functions on a disjoint union of $s$ circles. It is clear that for
every $V\in Gr(H^s)$  obtained by means of Krichever construction the
stabilizer $Stab_V$ of $V$ in $({\rm Lie }\ \Gamma)^s$ consists of all
functions defined on the union of circles $\partial D_1,...,\partial
D_s$ and admitting a  holomorphic continuation to $\Sigma \setminus
(D_1\cup ...\cup D_s)$. One can check that $Stab _V$ belongs to
$Gr(({\rm Lie}\ \Gamma )^s)$ for an appropriate definition of
semi-infinite structure in ${\rm Lie}\ \Gamma$. (We can consider
${\rm Lie}\ \Gamma$ as a subset of $H=L^2(S^1)$; the standard
semi-infinite structure on $L^2(S^1)$ induces a semi-infinite 
structure on pre Hilbert space Lie $\Gamma$.) We see that the 
locus  $P_s({\rm Lie}\ \Gamma)$ contains all points of 
$Gr(H^s)$ obtained by Krichever construction (Krichever
locus). Moreover, one can prove that Krichever locus coincides 
with $P_s({\rm Lie}\ \Gamma)$. (For $s=1$ this follows from [2]; 
see also [1]. The case $s>1$ can be treated in similar way; see 
the paper [3], devoted to various generalizations of the above 
statement.) 

  Let us describe an interesting duality transformation of $Gr(H^n)$. 
We identify $H$ with $L^2(S^1)$ equipped with standard semi-infinite 
structure. Then we can introduce bilinear inner product in $H$ by 
means of the formula 

$$(f,g)=\int _{|z|=1}f(z)g(z)dz.$$

This inner product induces inner product in $H^m$: if $f=(f_1,...,f_m) 
\in H^m,\  g=(g_1,...,g_m)\in H^m$, then 

$$(f,g)=\sum _k\int _{|z|=1}f_k(z)g_k(z)dz.$$

Let us define a linear operator $L:H\rightarrow H$ by the formula 

$$(Lf)(z)=f(-z).$$

  The map $L^m:H^m\rightarrow H^m$ transforms $(f_1,...,f_m)$ into $(Lf_1,...,Lf_m)$. The symbol $V^{\perp}$ denotes the orthogonal 
complement of $V\subset H^m$ with respect to bilinear inner product; 
the symbol $\lambda (V)$ stands for $L^mV^{\perp}$. 

  {\bf Theorem 3.} If $V\in Gr^{(k)}(H^m)$ then $V^{\perp} \in 
Gr^{(-k)}(H^m)$ and $\lambda (V) \in Gr^{(-k)}(H^m)$. The maps 
$\lambda :Gr(H^m)\rightarrow Gr(H^m)$ commute with
maps $\nu _{m,n}$ and $\sigma ^{(m)}$; in particular 

$$\lambda \circ \sigma ^{(m)} =\sigma ^{(m)}\circ \lambda.$$

  To prove the first statement we represent the space $V\subset 
H$ as an image of operator $\alpha :H_+\rightarrow H$, where 
$\alpha (h) =(Fh,Bh),\  F:H_+\rightarrow H_+$ is Fredholm, 
and $B:H_+\rightarrow H_-$ is compact. (We restrict ourselves 
to the case $m=1$.) Then orthogonal complement $V^{\prime}$ to 
the space $V$ with respect to hermitian inner product $<f,g>=
\int _0^{2\pi} f(\varphi )\bar {g(\varphi)}d\varphi$ consists 
of pairs $(h_+,h_-)$ obeying the equation $F^* h_+=B^*h_-=0$. 
(Here $*$ denotes Hermitian conjugation.) If $V=H_+$ then 
$V^{\prime}=H_-$; this means that we should expect that in the 
case when $V$ is close to $H_+$ the space $V^{\prime}$ should 
be close to $H_-$. In other words, we should expect that for 
$V\in Gr(H)$ the space $V^{\prime}$ belongs to $Gr(H)$ defined 
by means of semi-infinite structure when the roles of $H_+$ 
and $H_-$ are interchanged. This fact immediately follows 
from Lemma 1. Using that 

$$V^{\perp}=z\bar {V}^{\prime},\  \bar {H}_-=z^{-1}H_+,$$

we obtain that $V^{\perp}$ belongs to $Gr(H)$
original semi-infinite structure. It is easy to check that 
$L\in GL(H)$, therefore $LV^{\perp}$ also belongs to $Gr(H)$. 
Using Lemma 1 one can also calculate the index of $V^{\perp}$ 
and $LV^{\perp}$. 
   
  It is evident that $\lambda$ commutes with $\nu _{m,n}$; let 
us prove that $\lambda$ commutes with $\sigma ^{(m)}$. We start 
with a remark that 

  $$\sigma ^{(m)}(V)=\pi(V\cap R_K)$$

where $R_K$ denotes the subspace of $H^m$, that consists of 
points $(f_1,...,f_m)\in H^m$ obeying $f^{m-1}=Kf_m$, and 
$\pi$ stands for natural projection of $H^m$ onto $H^{m-2}$ (i.e. $\pi(f_1,...,f_m)=(f_1,...,f_{m-2}$.) It is easy to check that 

 $$\sigma ^{(m)}(V)^{\perp}=(\pi ^T)^{-1}((V\cap R_K)^{\perp})=
(\pi ^T)^{-1}(V^{\perp}+R_K^{\perp}).$$

 Here $\pi ^T :H^{m-2}\rightarrow H^m$ is an operator, adjoint 
to $\pi$ with respect to bilinear inner product; it is easy to 
see that it transforms $(f_1,...,f_{m-2})$ into 
$(f_1,...,f_{m-2},0,0)$. The space $R_K^{\perp}$ consists of points $(g_1,...,g_m)\in H^m$ obeying $g_{m-1}=-Kf_m$. Using these facts we 
obtain that $(f_1,...,f_{m-2})\in \sigma ^{(m)}(V)^{\perp}$ if there 
exists a point $(f_1,...,f_{m-2},f_{m-1},f_m)\in V^{\perp}$ 
satisfying $f_{m-1}=-Kf_m$. In other words, 

 $$\sigma ^{(m)}(V)^{\perp}=\sigma _{-K}^{(m)}(V^{\perp})$$

(The subscript $-K$ means that we replace the involution $K$ in 
the definition of $\sigma ^{(m)}$ with the involution $-K$). To 
check that $\lambda$ commutes with $\sigma ^{(m)}=\sigma _K^{(m)}$ 
we notice that 

$$\lambda (\sigma_K^{(m)}(V))=L^m(\sigma_K^{(m)}(V)^{\perp})=L^m\sigma _{-K}^{(m)}(V^{\perp}),$$

$$\sigma_K^{(m)}(\lambda (V))=\sigma _K^{(m)}(L^mV^{\perp})=L^m\sigma _{L^{-1}KL}^{(m)}(V^{\perp}).$$

It remains to take into account that $L^{-1}KL=-K$.
  
{\bf 2. Generalized moduli spaces. (Modular operads.)}
\vskip .1in
  One can reformulate the above results in the following way. 
Let us consider a sequence of $S_m$-spaces $P_m$. Here $S_m$ 
denotes the symmetric group , i.e. a group of permutations 
of $m$ elements $\{1,2,...,m\}$. $S_m$-space is by definition 
a topological space with left action of the group $S_m$. The 
group $S_k$ for $k<m$ is embedded into the group $S_m$ as a 
subgroup consisting of permutations leaving intact $k+1,...,m$. 
The group $S_m\times S_n$ is naturally embedded in $S_{m+n}$; 
namely, $S_m$ permutes first $m$ indices and  $S_n$ permutes 
last $n$ indices. Let us fix maps $\nu _{m,n}:P_m\times 
P_n\rightarrow P_{m+n}$ and $\sigma ^{(m)}:P_m\rightarrow 
P_{m-2}$. We will say that these data specify a MO if 
the following conditions are satisfied:

  0. The maps $\nu _{m,n}$ determine an associative multiplication 
in $\cup P_m$.

  1. $\nu _{m,n}\circ (\rho\times\tau)=(\rho\times \tau)\circ
\nu_{m,n}$ for every $\rho \in S_m,\  \tau \in S_m$,

  2. $\sigma ^{(m)}\circ \nu=\nu\circ \sigma ^{(m)}$ for every $\nu\in
S_{m-2}$,
 
  3. $\sigma ^{(m)}\circ \lambda =\sigma^{(m)}$ if $\lambda =(m,m-1)$
(i.e.  $\lambda$ permutes two last indices),
 
  4. $\sigma ^{(m)}$ commutes with $\sigma ^{(m)}\cdot \mu$ where $\mu
\in S_m$ is a permutation obeying $\mu (m-1)\leq m-2,\  \mu(m)\leq
m-2$,
 
  5. $\sigma ^{(n)}\circ \nu_{m,n}=\nu_{m,n-2}\circ \sigma ^{(n)}$,

  6. $\nu_{n,m}\circ \alpha =\beta \circ \nu _{m,n}$,
where $\alpha$ stands for the natural map $P_m\times P_n\rightarrow
P_n\times P_m$ (transposition), $\beta$ denotes a permutation
$(1,...,m+n)\rightarrow (m+1,...,m+n,1,...,m)$.

   The notion of MO is almost equivalent to the notion of modular
operad introduced in [4], as was pointed out to me by A. Voronov. 
The definition given in [4] is very complicated (in particular it 
is based on the notion of operad, which is not so simple by itself). 
I believe that the notion introduced above is simpler and more 
fundamental than the notion of operad, therefore it is unreasonable 
to use the term "operad" in its name. However, I did not find any 
good word for this notion.
  
   Notice, that the conditions 2),3) permit us to define maps $\sigma
_{a,b}^{(m)}:P_m\rightarrow P_{m-2}$ satisfying

   $$\sigma _{a,b}^{(m)} \circ \lambda =\sigma _{\lambda(a),\lambda
(b)}^{(m)},\ \ \sigma _{a,b}^{(m)} =\sigma _{b,a}^{(m)},\ \
\sigma_{m-1,m}^{(m)}=\sigma ^{(m)}$$

(Here $a,b\in \{ 1,...,m\} ,a\not= b,\lambda \in S_m$.) Using the
maps $\sigma _{a,b}^{(m)}$ we can rewrite 4) in the form: $\sigma
_{a,b}^{(m)}\circ \sigma _{a^{\prime},b^{\prime}}^{(m)}=\sigma
_{  a,b}^{(m)}\circ\sigma _{a,b}^{(m)}$ if $a,b,a^{\prime},b^{\prime}$
are distinct.
 
  The symmetric group $S_m$ acts naturally on the spaces $Gr_m=Gr(H^m)$
and $P_m({\cal G})$ described above. The maps $Gr(H^m)\rightarrow
Gr(H^{m-2})$ and $P_m({\cal G})\rightarrow P_{m-2}({\cal G})$ were
constructed in Theorem 3. One can define $\nu _{m,n}$ as a map
transforming a pair of subspaces $V\subset H^m,\  V^{\prime}\subset
H^n$ into subspace $V\oplus V^{\prime}\subset H^{m+n}=H^m\oplus H^n$;
it maps $Gr (H^m)\times Gr (H^n)$ into $Gr (H^{m+n})$ and $P_m({\cal
G})\times P_n({\cal G})$ into $P_{m+n}({\cal G})$. It is easy to check
that the conditions 0)-6) are safisfied. We obtain

  {\bf Theorem 4.} The $S_m$-spaces $Gr(H^m)$ together with maps 
$\nu_{m,n}: Gr(H^m)\times Gr(H^n)\rightarrow Gr(H^{m+n})$ and 
$\sigma ^{(m)}: Gr(H^m)\rightarrow Gr(H^{m-2})$ constitute a MO. 
Similar statement is true for the $S_m$-spaces $P_m({\cal G})$.

  We mentioned already that it is not necessary to assume that 
the operator $K$ in the definition of semi-infinite structure 
on $H$ is an involution. It is possible to construct the maps 
$\nu _{m,n}:Gr(H^m)\times Gr(H^n)\rightarrow Gr(H^{m+n})$ and  
$\sigma ^{(m)}:Gr(H^m)\rightarrow Gr (H^{m-2})$ without this 
assumption and to prove that all conditions in the definition 
of MO, except condition 3), are satisfied. This statement 
remains correct if we replace $Gr(H^m)$ with $P_m({\cal G})$.
  
  Let us assume that the $S_m$-spaces $P_m$ have an $S_m$-invariant decomposition $P_m=\cup _kP_m^{(k)}$ and that the maps $\nu _{m,n}$ 
and $\sigma ^{(m)}$ respect this decomposition (more precisely, 
$\nu _{m,n}$ maps $P_m^{(k)}\times P_n^{(l)}$ into $P_{m+n}^{(k+l)}$ 
and $\sigma ^{(m)}$ maps $P_m^{(k)}$ into $P_{m-2}^{(k)}$). Then we 
will talk about graded MOs. It is easy to see that the MO described 
in Theorem 4 can be considered as graded MOs with respect to 
decompositions $Gr(H^m)=\cup Gr^{(k)}(H^m),\ \  P_m({\cal G})=\cup P_m^{(k)}({\cal G})$.

 Notice that for every MO  $(P_n,\nu _{m,n},\sigma ^{(n)})$ one 
can define maps $P_m\times P_n\rightarrow P_{m+n-2}$ taking 
composition $\sigma ^{(m+n)}\circ \nu_{m,n}$. In particular, 
for $m=n=2$ we obtain a map $P_2\times P_2\rightarrow P_2$; 
this map determines a structure of semigroup on $P_2$. If 
$P_n$ is a graded MO ($P_n=\cup P_{\chi ,n}$) we obtain a 
structure of semigroup on $P_{0,2}$. (For the MO constructed 
by means of moduli spaces of complex curves $P_{0,2}$ is 
called Neretin semigroup.) If we have an algebra $(P_n,E,
\alpha _n)$ over $P_n$ then under certain regularity 
conditions the map $\alpha _2$ determines a representation 
of the Lie algebra of semigroup $P_{0,2}$ (or $P_2$) in 
the space $E$. Of course, not always we can speak about
Lie algebra of a semigroup, but such a Lie algebra exists 
in many interesting cases. In particular, the Lie algebra 
of Neretin semigroup coincides with complexified Lie algebra 
of diffeomorphism group of a circle.

  MOs defined above should be called topological MOs. One can 
define linear MOs regarding $P_m$ as $S_m$-modules. (Then we 
should consider $\nu _{m,n}$ as a bilinear map, i. e. as a 
linear map of tensor product $P_m\otimes P_n$ into $P_{m+n}$, 
and $\sigma^{(m)}$ as a linear map $P_m\rightarrow P_{m-2}$. 
In the definition of graded MO we should replace disjoint 
union with direct sum.) It is easy to see that homology 
groups of topological MO constitute a linear MO. A simple, 
but very important, example of linear MO (standard linear 
MO) can be defined if we have a linear space $E$ equipped 
with symmetric bilinear inner product. Then we can take 
the $m$-th tensor degree $E^{\otimes m}$ as $P_m$. The 
definitions of maps $\nu _{m,n}: E^{\otimes m}\otimes 
E^{\otimes n}\rightarrow E^{\otimes (m+n)}$ and 
$\sigma^{(m)}:E^{\otimes m}\rightarrow  E^{\otimes (m-2)}$ 
are obvious. If bilinear inner product is not symmetric 
the maps $\nu_{m,n}$ and $\sigma ^{(m)}$ obey all conditions 
in the definition of MO, except 3). 
 
  A homomorphism of MO $P_m$ into a MO $P_m^{\prime}$ can be 
defined as a collection of maps $\alpha _m:P_m\rightarrow 
P_m^{\prime}$, commuting with the operations $\nu_{m,n},
\sigma ^{(m)}$. A homomorphism of MO into itself is called 
an automorphism. Theorem 3 can be interpreted as a statement 
that the maps $\lambda :Gr(H^m)\rightarrow Gr(H^m)$ constitute 
an automorphism of the MO $Gr(H^m)$. Under certain conditions 
one can check that $\lambda$ induces also an automorphism of 
MO $P_m({\cal G})$. (One should assume that for every $\gamma 
\in \alpha ({\cal G})$ we have $\gamma ^T\in \alpha ({\cal G})$ 
and $L\gamma L^{-1}\in \alpha ({\cal G})$. Here $\alpha ({\cal G})$ 
stands for the image of ${\cal G}$ by the embedding $\alpha :{\cal G} 
\rightarrow gl(H),\  \gamma ^T$ denotes an operator adjoint to 
$\gamma$ with respect to bilinear inner product in $H=L^2(S^1),\  (Lf)(z)=f(-z)$.) 

  Let us define an algebra over (topological) MO as a homomorphism 
of it into linear MO described above. In other words, an algebra 
over MO $P_n$ is a collection of maps $\alpha _m :P_m\rightarrow 
E^{\otimes m}$ such that

  $$\sigma ^{(m)}\alpha _m=\alpha _{m-2}\sigma ^{(m)},$$

  $$\nu_{m,n}(\alpha_m\times \alpha _n)=\alpha _{m+n}\nu_{m,n}.$$

  We will assume that $\alpha _m(x),\  x\in P_m$, is defined 
only up to a constant factor. (One can say that we consider 
projective algebras; if $\alpha _m(x)$ is well defined we 
will talk about algebra in strict sense.) Then for appropriate 
choice of MO the notion of algebra over MO corresponds to the 
notion of conformal or superconformal field theory; if the 
maps $\alpha _n$ are defined uniquely the central charge of
corresponding conformal field theory vanishes.

We will not exclude the case when inner product in linear space
$E$ entering the definition of algebra over MO is determined only
on a subset $X$ of $ E^2$ ( then it is naturally to assume that
$E$ is a topological linear space and $X$ is dense in $E^2$ ).
If it is necessary to emphasize that we are dealing with this
case, we will use the term " generalized algebra ".

If one uses  Hilbert-Schmidt operators in the definition of 
Grassmannian, it is well known [1] that for every element 
$V\in Gr (H)$ one can construct an element $\Psi _V$ of the 
fermionic Fock space ${\cal F}(H)$, defined up to a factor. 
We assume that the space $H$ is equipped with antiunitary 
involution $f\rightarrow \bar {f}$, preserving $H_+$ and 
$H_-$. Then the Clifford algebra $Cl(H)$ can be defined 
as an associative unital algebra with generators $\psi (f),
\  \psi ^+(f)$, depending linearly of $f\in H$ and satisfying

$$[\psi (f),\psi ^+(f^{\prime})]_+=(f,f^{\prime}), [\psi^+(f),\psi^+(f^{\prime})]_+=[\psi(f),\psi(f^{\prime})]_+=0$$

 (Here ( , ) denotes bilinear inner product related with 
Hermitian inner product in $H$ by the formula $(f,g)=<f,\bar {g}>$.) 
Fock space ${\cal F}(H)$ can be defined as a space of representation 
of $Cl(H)$, that contains a cyclic vector $\Phi$,obeying $\psi 
(f)\Phi =0$ for $f\in H_+,\  \psi^+(f)\Phi =0$ for $f\in H_-$. 
(We assume that $\psi ^+(\bar {f})$ is Hermitian conjugate to 
$\psi (f)$.) If $V\in Gr (H)$ then $\Psi _V$ can be defined as a 
vector from ${\cal F}(H)$ satisfying the conditions $\psi (f) 
\Psi _V=0$ for $f\in V,\  \psi ^+(f)\Psi _V=0$ for $f\in V^{\bot}$. 
(Here $V^{\bot}$ stands for orthogonal complement to $V$ with 
respect to bilinear inner product.)

  Notice that for every vector $\Psi \in {\cal F}(H)$ one can 
define two orthogonal subspaces

  $$Ann \Psi =\{ f\in H|\psi (f)\Psi =0\}$$

and

   $$Ann^+\Psi =\{ f\in H|\psi ^+(f)\Psi =0\}$$

One can prove the following

  {\bf Lemma 2.} If $Ann \Psi \supset V,\  Ann ^+\Psi\supset 
KW,\  V\in Gr^{(k)}(H),\  W\in Gr^{(-k)}(H)$ then $\Psi 
=\Psi _V,\  V=Ann\Psi,\  W=V^{\bot}=K Ann^+\Psi$.

   We will define also a bilinear inner product (  , ) in ${\cal
F}={\cal F}(H)$, obeying the condition that $-\psi (Kf)$ is adjoint to
the operator $\psi (f)$ and $-\psi ^+(Kf)$ is adjoint to $\psi ^+(f)$
with respect to this inner product. Using this bilinear inner product
we define a linear MO ${\cal F}^n$.

  One can check the formula

\begin {equation}
\Psi_{\sigma ^{(n)}V}=\sigma ^{(n)}\Psi _V \ \  {\rm for }\  V\in
Gr(H^n)
\end {equation}
(We use the identification ${\cal F}(H^n)={\cal F}(H)^{\otimes n}$ 
in (5). This identification permits us to consider ${\cal F}(H^n)$ 
as a linear MO.) To prove (5) we represent  $\Psi _V\in {\cal F}^
{\otimes n}$ in the form

$$\Psi _V=\sum \alpha _i\otimes A_i\otimes B_i$$

where $\alpha _i\in {\cal F} ^{\otimes (n-2)},\  A_i\in {\cal F},
\  B_i\in {\cal F}$. By definition for every $(f_1,...,f_n)\in V$ 
we have

\begin {equation}
\sum (\psi (f)\alpha _i)\otimes A_i\otimes B_i+\sum \alpha _i\otimes
(\psi (f_{n-1})A_i)\otimes B_i+\sum \alpha _i\otimes A_i\otimes (\psi
(f_n)B_i)=0
\end {equation}
where $\psi (f)=\psi (f_1)\otimes 1\otimes ...\otimes 1+1\otimes
\psi(f_2)\otimes 1\otimes ...\otimes 1+1\otimes ...\otimes 1\otimes\psi
(f_{n-2}).$ Let $(f_1,...,f_{n-2})\in \sigma ^{(n)}V$. Then it can be
obtained from a point $(f_1,...,f_{n-2},f_{n-1},f_n)\in V$ with
$f_{n-1}=Kf_n$. Applying operator $\sigma^{(n)}$ to $\Psi _V$ we obtain

$$\sigma ^{(n)}\Psi_V=\sum \alpha _i(A_i,B_i).$$

Now we can apply $\sigma ^{(n)}$ to (6). Using the relation

$$\sum \alpha _i(\psi(f_{n-1})A_i,B_i)+\sum \alpha _i(A_i,\psi
(f_n)B_i)=0$$

for $f_{n-1}=Kf_n$ we see that

$$\psi (f)\sigma ^{(n)}\Psi_V=0$$

for every $f=(f_1,...,f_{n-2})\in \sigma ^{(n-2)}V$. In other words,

$$\sigma ^{(n)}V\subset Ann\sigma^{(n)}\Psi _V$$

In similar way we prove

$$K\sigma ^{(n)}(KV^{\bot})\subset Ann^+\sigma ^{(n)}\Psi_V.$$

Taking into account that

$${\rm index}\  \sigma^{(n)}V+{\rm index}\  \sigma ^{(n)}(KV^{\bot})=0$$

and using Lemma 2 we obtain (5).

The following statement follows immediately from (5).

{\bf Theorem 5.} The sequence of maps

$$V\in Gr(H^m)\rightarrow \Psi _V \in {\cal F}(H^m)=
{\cal F}^{\otimes m}$$

determines an algebra over a MO $Gr(H^m)$.

 It is important to emphasize that in all definitions and 
theorems above one can replace spaces with superspaces; only 
minor modifications are required. (The only exception is 
Theorem 5, that requires more essential modification; see 
Appendix.) In particular, one can consider super Grassmannian. 
To give an example of a superspace with semi-infinite structure 
we can consider the space $H=H^{m|n}$ of functions on $S^1$ 
taking values in $(m|n)$-dimensional linear complex superspace 
${\bf C}^{m|n}$. The  definitions of $H_+,\  H_-$ and $K$ repeat 
definitions for $H=L^2(S^1)$. Notice, that $H^{1|1}$ can be considered 
also as  the space of functions on the supercircle, i.e. as the space 
of functions $F(z,\theta)=f(z)+\varphi (z)\theta ,\  |z|=1,\ \theta $ 
is an odd variable, $f(z)$ is an even function on $S^1$ and $\theta (z)$ 
is an odd function on $S^1$.

 The following construction will play later an important 
role. For every manifold $M$ we define a supermanifold 
$\tilde {M}=\Pi TM$ as the space of tangent bundle with 
reversed parity of fibers. One can define an odd vector 
field $\hat {Q}=\xi ^i{\partial \over \partial x^i}$ on
$\tilde {M}$ (here $x^i$ are coordinates on $M,\  \xi^i$ 
are odd coordinates in tangent spaces.) Functions on $M$ 
can be identified with differential forms on $M$, then 
the operator $\hat {Q}$ corresponds to exterior differential. 
Notice, that $\{ Q,Q\} =0$; this means that $M$ is a $Q$-manifold 
in the terminology of [6]. One can construct the supermanifold 
$\tilde {M}$ also in the case when $M$ is a supermanifold. 
Then again differential forms on $\tilde {M}$ can be considered 
as functions on $M$, however not all functions on $\tilde {M}$ 
correspond to differential forms. One can describe $\tilde {M}$
also as the space of maps of $(0|1)$-dimensional superspace ${\bf
R}^{0|1}$ into $M$. It is easy to see that this construction is
functorial: for every map $f:M\rightarrow M^{\prime}$ one can 
define naturally a map $\tilde {f}:\tilde {M}\rightarrow 
\tilde {M}^{\prime}$. It follows from this remark that in the 
case when $M$ is a (super) Lie group $\tilde {M}$ is a (super) 
Lie group (the multiplication map $M\times M\rightarrow M$ 
generates multiplication  $\tilde {M}\times \tilde {M} 
\rightarrow \tilde {M}$). If $M$ is a Lie algebra then
$\tilde {M}$ is a (super) Lie algebra.(If $l_n$ are 
generators of $M$, $f_{jk}^i$ are corresponding structure 
constants, then $\tilde {M}$ has even generators $l_n$ and odd 
generators $b_n$ with commutation relations: $[l_m,l_n]=f_{mn}^kl_k,
\  [l_n,b_m]=f_{mn}^kb_k,\ [b_m,b_n]_+=0$.)
 
  If a Lie group $G$ acts on $M$ then $\tilde {G}$ acts on 
$\tilde {M}$, transitive action on $G$ induces transitive 
action of $\tilde {G}$. If  $M=G/G_0$ then $\tilde {M}$ 
can be identified with $\tilde {G}/\tilde {G}_0$.

  The supermanifold $\tilde {M}=\Pi TM$ is equipped with 
natural volume element $dV=\Pi dx^id\xi ^i$ (this volume 
element does not depend on the choice of coordinates on 
the (super)manifold $M$). Therefore we can integrate a 
function on $\tilde {M}$ over $\tilde {M}$ or over $\tilde {L}$ 
where $L$ is a submanifold of $M$. (Of course, we should make some
assumptions about behavior of the function at infinity to guarantee the
convergence of the integral). If $M$ is an ordinary manifold then the
theory of integration of functions on $\tilde {M}$ is equivalent to the
theory of integration of differential forms on $M$. However, if $M$ is
a supermanifold then the functions on $\tilde {M}$, corresponding
 to differential forms on $M$, are not integrable. (By definition, such 
functions depend polynomially on $\xi ^i$. If some of coordinates 
$x^i$ are odd, corresponding $\xi ^i$ are even and the function 
does not decrease at infinity.) It is necessary to mention that 
we can integrate also generalized functions (distributions) over
$\tilde {M}$.

  The  volume element $dV $ on $\tilde {M}$ is $Q$-invariant. This 
means, that

$$\int _{\tilde {L}}(\hat {Q} f)dV=0$$

for every function $f$ on $\tilde {M}$ and every submanifold 
$L\subset M$. It is easy to check that for $Q$-invariant 
function $\varphi$ on $\tilde {M}$ the integral

$$\int _{\tilde {L}}\varphi dV$$

does not change by continuous deformation of the 
submanifold $L$. (For the case when $M$ is an ordinary 
manifold one can derive this fact from
the remark that $Q$-invariant function $\varphi$ on 
$\tilde {M}$ can be considered as a closed differential 
form on $M$.)

 If   $H=L^2(S^1)$ then $\tilde {H}$ can be regarded as 
$H^{1|1}$ (as the space of functions on the supercircle). 
This remark permits us to embed $\tilde {Gr}(H)$ into 
$Gr (H^{1|1})$. One can consider for example
$\tilde {Gr}^{(k)}(H)$ as a homogeneous space where 
$\tilde {GL}(H)$ acts transitively and utilize the 
fact the $\tilde {GL}(H)$ acts on $\tilde {H}=H^{1|1}$ 
and therefore on $Gr(H^{1|1})$. However, it is useful 
to describe the embedding $\tilde {Gr}(H)$ into 
$Gr(H^{1|1})$ explicitly. An arbitrary point $W\in 
Gr(H^{1|1})$ can be specified by means of a Fredholm
 operator $A:H^{1|1}_+\rightarrow H^{1|1}_+$ and a compact
 operator $B:H^{1|1}_+\rightarrow H^{1|1}_-$. These operators can be 
written as $2\times 2$-matrices

 $$\left( \begin {array}{cc}
  A_{11}&A_{12}\\
  A_{21}&A_{22}
 \end {array} \right) ,\ \ \
 \left( \begin {array}{cc}
  B_{11}&B_{12}\\
  B_{21}&B_{22}
 \end {array} \right) $$

where $A_{\alpha \beta}:H_+\rightarrow H_+,\  
B_{\alpha \beta }:H_+\rightarrow H_-$, diagonal entries are 
even, off-diagonal entries are odd. One can check that $W$ 
belongs to $\tilde {Gr}(H)\subset Gr (H^{1|1})$ if it can be 
represented by means of operators $A$ and $B$ with matrices 
obeying $A_{11}=A_{22},\  A_{12}=0,\  B_{11}=B_{22},\  B_{12}=0$.

 {\bf 3. Generalized string backgrounds.}
 \vskip .1in
   To calculate string amplitudes corresponding to a 
conformal field theory with critical central charge we 
should "add ghosts" to obtain topological conformal 
field theory with vanishing central charge. Such a TCFT 
can be considered as "string background". In other words 
one can define corresponding string amplitudes, even 
in the case when the TCFT does not correspond to any CFT 
(matter and ghosts are not separated.) In this section 
we will describe an abstract analog of these constructions.

 We say that a MO $(P_n,\nu _{m,n},\sigma ^{(m)})$ is a 
$Q$ -MO if every space $P_n$ is a $Q$-manifold (i. e. $P_n$ 
is a supermanifold equipped with a vector field $Q=Q^{(n)}$ 
obeying $\{ Q,Q\}=0$) and maps $\nu_{m,n},\sigma ^{(m)}$ 
are compatible with $Q$-structures on $P_n$ (for example, 
$\sigma _*^{(m)} Q^{(m)}=Q^{(m-2)}$). A $Q$-homomorphism of 
$Q$-MO $P_n$ into a $Q$-MO $P_n^{\prime}$ is defined as a 
collection of maps $\rho _n:P_n\rightarrow  P_n^{\prime}$
that determine a homomorphism of MOs and are compatible with
$Q$-structures on $P_n, P_n^{\prime}$ (i. e. $\rho _n$ 
transforms the vector field $Q$ on $P_n$ into corresponding 
vector field on $P_n^{\prime}$). If we have a linear MO $P_n$ 
we can introduce a notion of linear $Q$-structure on it, 
requiring that all vector fields $Q^{(n)}$ are linear, i. e. 
have the form

 $$(Q^{(n)})^a=(Q^{(n)})_b^az^b$$

where $( Q^{(n)})_b^a$ is a matrix of parity reversing linear 
operator, having zero square (such an operator is called a 
differential). In paricular, if $E$ is a linear superspace, 
equipped with a differential $d$ and $E$ is equipped with 
$d$-invariant inner product (i. e. $d=-d^*$) then one can 
construct a linear $Q$-MO with spaces $P_n=E^{\otimes n}$. 
(The differential $d$ on $E$ determines a differential $d: 
E^{\otimes n}\rightarrow E^{\otimes n}$) A linear $Q$-MO 
$E^{\otimes n}$ constructed this way will be called a 
standard  linear $Q$-MO.

 We define a $Q$-algebra $(P_n,E,\alpha _n)$ over a $Q$-MO 
$P_n$ as a $Q$-homomorphism of $Q$-MO $P_n$ into standard 
linear $Q$-MO $E^{\otimes n}$. Let us emphasize that although 
we assumed that the maps $\alpha _n$ are defined only up to a 
factor when we considered an algebra over MO, in the definition 
of $Q$-algebra we assume that $\alpha _n$ are well defined (i.e. 
$Q$-algebra should be an algebra in strict sense).

 Let us consider an arbitrary MO $P_n$. Then we can construct in 
natural way a $Q$-MO $\tilde {P}_n$. (Recall that $\tilde {P}_n = 
\Pi TP_n=\{R^{0|1}\rightarrow P_n\}$.) As was explained above this 
construction is functorial; hence the maps $\nu _{m,n},\sigma ^{(m)}$, specifying the MO $P_n$ induce maps $\tilde {\nu }_{m,n},\tilde {\sigma}^{(m)}$, that determine a MO $\tilde {P}_n$. We obtain a 
$Q$-MO this way because every $\tilde {P}_n$ has a natural 
$Q$-structure. A $Q$-algebra $(\tilde {P}_n,F,\beta _n)$ over 
$Q$-MO $\tilde{P}_n$ is called a $Q$-extension of an algebra 
$(P_n,E,\alpha _n)$ over a MO $P_n$ if there exists such a 
linear map $\rho :E\rightarrow F$ for every $x\in P_n\subset 
\tilde {P}_n,\ \  e\in E$, we have $(\rho(e),\beta _n(x))= 
(e,\alpha _n(x))$. If the algebra $(P_n,E,\alpha _n)$ corresponds 
to conformal field theory then we can obtain its $Q$-extension 
adding ghosts to the matter sector of the theory.

  Let us consider a MO  $(P_n,\nu _{m,n},\sigma ^{(n)})$. We say 
that this MO is an EMO (equivariant modular operad) if for every 
$m$ the group $(S^1)^m$ acts on $P_m$; this action should be 
compatible with $S_m$-action and with maps $\nu _{m,n},
\sigma ^{(m)}$. More precisely, we assume that

 1. $s\circ g\circ s^{-1}=s(g)$ where $g\in (S^1)^m,\  s\in S_m,\
g\rightarrow s(g)$ denotes the natural action of $S_m$ onto $(S^1)^m$.

 2. $\nu _{m,n}\circ (g\times g^{\prime})=(g\times g^{\prime})\circ
\nu _{m,n}$ for $g\in (S^1)^m,\  g^{\prime}\in (S^1)^n$.

 3. If $g=(g_1,...,g_m)\in (S^1)^m$ and $g_{m-1}=g_m^{-1}$ then $\sigma
^{(m)}\circ g=g^{\prime}\circ \sigma ^{(m)}$ where
$g^{\prime}=(g_1,...,g_{m-2})\in (S^1)^{m-2}$.

 We define $\rho ^{(m)}:P_m\times S^1\rightarrow P_{m-2}$ as a map
transforming a point $(x,\gamma) \in P_m\times S^1$ into a point 
$\sigma ^{(m)}(gx)$ where $g=(1,...,1,\gamma )\in (S^1)^m$. It is 
possible to generalize a notion of EMO, taking as a starting point 
the maps $\rho ^{(m)}$; we will not discuss this generalization here.

 The MO corresponding to conformal field theory can be considered as an
EMO. Recall that the elements of $P_m$ in this case can be considered
as surfaces with $m$ holes, having boundaries parametrized by the
circle $|z|=1$. The action of $(S^1)^m$ changes the parametrization of
boundary circles: $(z_1,...,z_m)\rightarrow (e^{i\alpha
_1}z_1,...,e^{i\alpha _m}z_m)$.

 Similarly, we can introduce a structure of EMO in the MO 
$Gr(H^k)$, assuming that $H$ has a semi-infinite structure 
$(H_+,H_-,K)$ and the group $S^1$ acts on $H$, in such a 
way that $H_+$ and $H_-$ are invariant subspaces and 
$K\circ g\circ K^{-1}=g^{-1}$ for every $g\in S^1$. If the 
semi-infinite structure on $H=L^2(S^1)$ is chosen in
standard way ($H_-$ spanned by $z^n,\  n\geq 0,\  H_+$ by 
$z^n,\  n<0,\ (Kf)(z)=z^{-1}f(z^{-1})$), then one can define 
an action of the group $S^1$ by the formula $g_{\alpha}(z)= 
e^{i\alpha }g(e^{2i\alpha}z)$.

 As we know for every MO $P_n$ the spaces $\tilde {P}_n$ 
constitute a $Q$-MO. We define a $Q$-EMO as  a $Q$-MO with 
action of $(\tilde {S}^1)^n$ satisfying natural conditions. 
It is easy to see that if $P_n$ constitute  an EMO the spaces 
$\tilde {P}_n$ constitute  a $Q$-EMO.

 Every linear space $F$ with inner product and $S^1$-action 
preserving this inner product determines a  linear EMO. A  
linear superspace with inner product and  linear operators 
$Q,l,b$,obeying $Q^2=0,\  b^2=0,\ l=[Q,b]_+$ determines a 
linear $Q$-EMO. (Operators $Q$ and $b$ should be parity 
reversing. Inner product should be invariant with respect 
to $Q,l,b$. We assume that $l$ generates an action of $S^1$.)

 One can construct a BV-algebra corresponding to an EMO. (This
construction is similar to the construction of [7].) Let us denote by
$C_n^k$ the  linear space of singular $k$-dimensional chains in $P_n$.
Using the map $\rho ^{(n)}:P_n\times S^1\rightarrow P_{n-2}$ we can
define a map $\Delta :C_n^k\rightarrow C_{n-2}^{k+1}$ by the formula
$\Delta (x)=\rho _*^{(n)}(x\times [S^1])$, where $[S^1]$ stands for the
fundamental cycle of $S^1$. The space $C_n=\sum _kC_n^k$ can be
considered as graded $S_n$-module; the $S_n$-invariant part of $C_n$
will be denoted by $C_k^{inv}$. It is easy to check that $\Delta$ is a
parity reversing operator acting from $C_n^{inv}$ into  $C_{n-2}^{inv}$
and that $\Delta ^2=0$ on  $C_n^{inv}$. The map $\nu _{m,n}:P_m\times
P_n\rightarrow P_{n+m}$ generates a map $(\nu _{m,n})_*:C_m^k\times
C_n^l\rightarrow C_{m+n}^{k+l}$. If $x\in C_m^{inv},\  y\in C_n^{inv}$
we denote by $x\cdot y$ an element of  $C_{m+n}^{inv}$ obtained by
means of symmetrizatiom of $(\nu_{m,n})_*(x\times y)$. Let us consider
now a graded  linear  space $C^{inv}=\sum C_n^{inv}$ (grading in
$C^{inv}$ is induced by the grading in $C_n^{inv}$. One can prove that
$\Delta :C^{inv}\rightarrow C^{inv}$ is a second order derivation with
respect to multiplication $(x,y)\rightarrow x\cdot y$ and therefore
$C^{inv}$ can be considered as a BV-algebra.

{\bf  4. String amplitudes.}
\vskip .1in
Let us consider a linear space $E$ with inner product and with
$S^1$-action preserving this inner product. As we mentioned such a
space determines a   linear EMO. If $P_n$ is an EMO we can consider an
algebra $(P_n,E,\alpha _n)$ as a homomorphism of $P_n$ into $E^{\otimes
n}$. (We require that maps $\alpha _n:P_n\rightarrow E^{\otimes n}$
commute with $(S^1)^n$-action.)

  To define string amplitudes corresponding to the algebra $(P_n,E,\alpha
_n)$ we should construct at first a $Q$-extension $(\tilde
{P}_n,F,\beta _n)$ of $(P_n,E,\alpha _n)$ (we should "add the ghosts").
This means that we should construct a $Q$-algebra over $Q$-EMO $\tilde
{P}_n$,that can be considered as a $Q$-extension  of our original  
algebra. More precisely, we should construct a linear space $F$ 
with inner product and odd operators $Q,b$, respecting this inner product
and obeying $Q^2=b^2=0$. We assume that the operator $l=[Q,b]_+$
generates an action of $S^1$; then $l,b$ generate an action of $\tilde
{S}^1$ on $F$ and an action of $(\tilde {S}^1)^n$ on $F^n$. The
$Q$-algebra  $(\tilde {P}_n,F,\beta _n)$ is specified by the maps
$\beta _n:\tilde {P}_n\rightarrow F^{\otimes n}$ that are compatible
with the action of $Q$ and $(\tilde {S}^1)^n$ on $\tilde {P}$ and
$F^{\otimes n}$.

  Recall that we assumed that the maps $\alpha _n$ are defined only up to
a factor. However, we assume that the maps $\beta _n$ in  $(\tilde
{P}_n,F,\beta _n)$ are defined unambiguosly. As we mentioned the
passage to $Q$-extension corresponds to adding ghosts to conformal
field theory. Our assumption means that we consider critical theory
when adding ghosts gives zero central charge. Now we define string
amplitudes starting with equivariant $Q$-algebra $(\tilde
{P}_n,F,\beta_n)$. (Notice that we can forget about original algebra
$(P_n,E,\alpha _n)$ at this stage. This remark corresponds to well
known fact that string amplitudes can be defined also in the case when
matter and ghosts are not separated; in other words we can consider a
topological conformal field theory, that does not correspond to a
conformal field theory, as a string background.)

  The space $F$ can be interpreted as space of states in the string
theory. However, not all of them should be considered as physical
states. The physical states $A$ should satisfy the conditions $lA=0,\
bA=0,\  QA=0$ where $l$ and $b$ are generators of the group $\tilde {S}^1$ 
($l$ is even, $b$ is odd, $l=[b,Q]_+$, hence $lA=0$ follows from $bA=0,\  QA=0$).  Two physical states $A,A^{\prime}\in F$ should be considered equivalent if $A^{\prime}-A=QB$ where $B\in F,\  lB=0,\  bB=0$. In other 
words, the space of physical states can be identified with homology of 
operator $Q$ acting in the space

  $$F^{rel}=\{A\in F|lA=0,\  bA=0\}.$$

Now we can define the scattering amplitude for physical states
$A_1,...,A_n$ in the following way. Let us consider a function

\begin {equation}
\Phi _{A_1,...,A_n}(x)=(A_1\otimes ...\otimes A_n,\beta _n(x))
\end {equation}
on $\tilde {P}_n$. We assumed that the map $\beta _n:\tilde
{P}_n\rightarrow F^{\otimes n}$ respects the action of $(\tilde {S}^1)^n$
and $Q$ on $\tilde {P}_n$ and $F^{\otimes n}$. It follows immediately
from invariance of inner product and from relation $lA_i=0,\  bA_i=0,\  
QA_i=0$ that the function (7) is $(\tilde {S}^1)^n$-invariant and $Q$-invariant. This means that $\Phi _{A_1,...,A_n}(x)$ can be 
considered as a $Q$-invariant function on $\tilde {P}_n/(\tilde {S}^1)^n$. 
We define "string amplitude" as an integral of $\Phi _{A_1,...,A_n}(x)$ 
over $\tilde {M}_n$ where $M_n$ denotes the "fundamental cycle" of $P_n/(S^1)^n$. One should emphasize that $P_n/(S^1)^n$ is in general infinite-dimensional (super)manifold therefore the notion of "fundamental cycle" is ill-defined. However, say, in the case where $P_n$ is a space 
of complex curves with $n$ holes, the infinite-dimensional space 
$P_n/(S^1)^n$ is homotopy  equivalent to a disjoint union of 
finite-dimensional orbifolds and the  fundamental cycle $M_n$ of 
$P_n/(S^1)^n$ can be defined as a sum of corresponding fundamental 
cycles. It follows from $Q$-invariance of $\Phi _{A_1,...,A_n}(x)$ 
that the integral

$$\int _{\tilde {M}_n}\Phi _{A_1,...,A_n}(x)$$

depends only on homology class of $M_n$. This integral gives the 
standard expression for the bosonic string amplitudes. Similar 
considerations can be applied to the case when $P_n$ is a space 
of superconformal manifolds; we obtain an  expression for the 
fermionic string amplitudes in this case.

   The definition of  physical states given above is not completely 
general. One should define the space of physical states as the 
equivariant cohomology of $F$. More precisely, we should consider 
the subspace $F^{inv}[\Omega ]$ of polynomials of indeterminate 
$\Omega $ taking values in $F^{inv}=\{ x\in F|lx=0\}$. One can 
define a differential $d$, acting on $F^{inv}$, by the formula 
$d=Q-\Omega b$. Then equivariant cohomology of $F$ can be 
identified with cohomology of $d$, acting on $F^{inv}[\Omega ]$. 
It is clear that physical states in the old sense can be considered 
as physical states in the new sense. One can prove that both notions 
coincide if every element $x$ of $F$ obeying $bx=0$ can be represented 
in the form $x=by$ and $F$ splits into a direct sum of eigenspaces of 
$l$ [5]. If $A_1,...,A_n$ are equivariant cocycles (i.e. $A\in F^{inv},
\  dA=0$) we define $\Phi _{A_1,...,A_n}(x,\Omega _1,...,\Omega _n)$ 
by means of the same formula (7).

   It is easy to see that the function $\Phi =\Phi _{A_1,...,A_n}(x,
\Omega _1,...,\Omega _n)$ is $(S^1)^n$-invariant (i.e. $l_1\Phi=...= 
l_n\Phi=0$) and satisfies the condition

$$(Q-\sum \Omega _ib_i)\Phi =0.$$

  The function $\Phi$ on $\tilde {P}_n$ can be considered as an 
equivariant differential form on $P_n$; the conditions above mean 
that this form is equivariantly closed (with respect to the action 
of the group $(S^1)^n$); it determines therefore an equivariant 
cohomology class. In the case when $(S^1)^n$ acts freely on $P_n$ 
one can prove that equivariant cohomology of $P_n$ is isomorphic 
to cohomology of $P_n/(S^1)^n$; we obtain a  cohomology class of 
$P_n/(S^1)^n$ that can be used to define string amplitudes in the 
same way as above. Similar construction can be applied also in the 
case when $P_n$ is a supermanifold.

  Notice that in the definition of EMO and in considerations of 
present section one can replace the group $S^1$ with any other 
group $G$. (The condition 3 in the definition of EMO should be 
slightly modified. Namely, we replace the relation $g_{m-1}=g_m^{-1}$ 
with the relation $g_{m-1}=\rho(g_m)$ where $\rho$ is a map of $G$ 
into itself.) Notice, that MO $Gr(H^m)$ where $H=L^2(S^1)$ can be 
considered as EMO with respect to the action of the group $\Gamma$ 
of non-vanishing twice differentiable functions on $S^1$. We will 
argue that this EMO and similar EMO's should be related to 
non-perturbative string theory.
 
  Let us consider a collection of maps $\alpha _n:P_n\rightarrow 
E^{\otimes n}$, that determine an algebra over MO $P_n$. If $\Sigma 
\in P_1$ has an automorphism group $G$, then this automorphism group 
acts naturally on $E$. In many interesting cases one can extend the 
action of automorphism group $G$ to an action of $G^n$ onto $P_n$ to 
obtain an EMO and an algebra over this EMO. For example, if the algebra 
at hand corresponds to CFT, one can take as $\Sigma$ a standard disk 
and consider $G$ as a group $S^1$ of rotations of the disk. The space 
$P_n$ for the MO related to CFT consists of complex curves with $n$ 
embedded standard disks; the action of $(S^1)^n$ onto $P_n$ comes 
from automorphism groups of these disks.

   If $P_n$ constitute an EMO with respect to the group $G$, then 
$\tilde {P}_n$ constitute an EMO with respect to the group 
$\tilde {G}$. To define "string amplitudes" we can start with 
a $G$-equivariant algebra $(P_n,E,\alpha _n)$ over $P_n$. (We 
say that an algebra $(P_n,E,\alpha _n)$ is $G$-equivariant, if 
$G$ acts linearly on $E$, preserving the inner product, and 
the maps $\alpha _n:P_n\rightarrow E^{\otimes n}$ commute with 
the action of $G^n$.) We should extend $E$ to a $\tilde {G}$-module 
$F$ (to a linear superspace $F$ with linear action of the group 
$\tilde {G}$). At the level of Lie algebras such a module can be 
described by means of even operators $L_n$ and  odd operators 
$b_n$ obeying

\begin {equation}
[L_m,L_n]=f_{mn}^kL_k,\  [L_m,b_n]=f_{mn}^kb_k,\  [b_m,b_n]_+=0,
\end {equation}
where $f_{mn}^k$ are structure constants of Lie algebra ${\cal G}$ 
of $G$. One should assume also that $F$ is equipped with an odd 
differential $Q$ obeying

\begin {equation}
L_m=[Q,b_m]_+
\end {equation}
and with inner product which is $Q$-invariant and $\tilde {G}$-invariant.
As we mentioned, the Lie algebra $\tilde {{\cal G}}$ of the group 
$\tilde {G}$ has even generators $L_n$ and odd generators $b_n$ 
obeying (8). Adding to $L_n, b_n$ an odd generator $Q$ satisfying 
(9) and $[Q,Q]_+=0$ we obtain a Lie algebra that will be denoted 
by ${\cal G}^{\prime}$. Corresponding extension of the group 
$\tilde {G}$ will be denoted by $G^{\prime}$.

  A $Q$-extension $(\tilde {P}_n,F,\beta _n)$ of an algebra 
$(P_n,E,\alpha _n)$ over EMO $P_n$ should be considered as 
a $Q$-algebra, that respects the action of $\tilde {G}^n$ in 
$\tilde {P}_n$ and $F^{\otimes n}$. In other words $\tilde {P}_n$ 
should be considered as EMO with respect to the group $G^{\prime}$ 
and $(\tilde {P}_n, F, \beta _n)$ should be a $G^{\prime}$-equivariant 
algebra. The space ${\cal A}$ of physical states must be identified 
with equivariant cohomology of $F$ with respect to the action of $G^{\prime}$.

  To define equivariant cohomology of $F$ we begin with the space 
$F\otimes \Phi (\Pi \tilde {{\cal G}})$ where 
$\Phi (\Pi\tilde {{\cal G}})$ stands for a space of functions on 
$\Pi\tilde {{\cal G}}$. One can consider various functional spaces 
and obtain different versions of the notion of equivariant 
cohomology. The standard notion of equivariant cohomology 
corresponds to the space of polynomial functions on 
$\Pi \tilde {{\cal G}}$. If Lie algebra ${\cal G}$ is 
$m$-dimensional polynomial functions on $\Pi \tilde {{\cal G}}$ 
can be identified with polynomials of $m$ odd variables 
$\omega _1,...,\omega _n$ and $m$ even variables 
$\Omega _1,...\Omega _n$. The group $G^{\prime}$ acts 
naturally on ${\cal G}^{\prime}$; this action is linear 
and therefore determines a linear action of $G^{\prime}$ 
on $\Pi{\cal G}^{\prime}$. One can identify functions on 
$\Pi \tilde {{\cal G}}$ with homogeneous functions on 
$\Pi{\cal G}^{\prime}\setminus \Pi \tilde {{\cal G}}$; 
using this identification we obtain an action of $G^{\prime}$ 
on $\Phi (\Pi\tilde {{\cal G}})$. (One can say also that we 
define the action of $G^{\prime}$ on $\Phi (\Pi\tilde {{\cal G}})$ 
using the embedding of $\Pi\tilde {{\cal G}}$ into projective space corresponding to linear space $\Pi {\cal G}^{\prime}$.) Combining $G^{\prime}$-action on $F$ and $G^{\prime}$-action on 
$\Phi (\Pi\tilde {{\cal G}})$ we obtain $G^{\prime}$-action on 
$F\otimes \Phi(\Pi\tilde {{\cal G}})$. In other words, we have 
$\tilde {G}$-action on $F\otimes \Phi(\Pi\tilde {{\cal G}})$ and 
a differential $Q_{tot}$ on this space.

   The differential $Q_{tot}$ acts on the set $(F\otimes \Phi (\Pi 
\tilde {{\cal G}})) ^{inv}$ of $\tilde {{\cal G}}$-invariant elements 
of $F\otimes \Phi (\Pi \tilde {{\cal G}})$; we define equavariant 
cohomology $H_{G}(F)$ of $F$ as cohomology of $Q_{tot}$ acting on 
$(F\otimes \Phi (\Pi \tilde {{\cal G}}))^{inv}$. If $G$ is an 
$m$-dimensional connected Lie group we represent an element of 
$F\otimes \Phi (\Pi \tilde {{\cal G}})$ as an $F$-valued function 
$\varphi (\omega _1,...,\omega _m,\Omega _1,...,\Omega _m)$. The 
condition of $\tilde {G}$-invariance means that

\begin {equation}
L_i\varphi +(\omega _{\alpha}f_{i\alpha}^k{\partial \over \partial \omega _k}+\Omega _{\alpha}f_{i\alpha}^k{\partial \over \partial \Omega _k})\varphi =0
\end {equation}

\begin {equation}
b_i\varphi +({\partial \over \partial \omega _i}+\omega _{\alpha}f_{i\alpha}^k{\partial \over \partial \omega _k})\varphi =0
\end {equation}

The second of these equations can be used to eliminate 
$\omega _1,...,\omega _m$ and to obtain a generalization of 
the definition of $S^1$-equivariant cohomology that we gave 
above. More precisely, we assign to every solution 
$\varphi (\omega_1,...,\omega _m,\Omega _1,..., \Omega _m)$ 
of (11) a function $\varphi (0,...,0,\Omega _1,...,\Omega  _m)$; 
we obtain one-to-one correspondence between solutions of (11) 
and functions depending on $\Omega_1,...,\Omega _m$. Using 
this correspondence we can define equivariant cohomology as 
cohomology of the operator $Q-\Omega ^ib_i$ acting on the 
space of $G$-invariant functions of $\Omega_1,...,\Omega _m$ 
taking values in $F$. As we mentioned one can modify the 
definition of equivariant cohomology considering various 
spaces of functions on $\Pi\tilde {{\cal G}}$. If 
$\Phi (\Pi {\cal G})$ stands for the space of all smooth 
functions corresponding equivariant cohomology of $F$ is 
denoted by $H_{G}^{\infty}(F)$ (more precisely, one should 
consider the space of smooth $F$-valued functions on 
$\Pi\tilde {{\cal G}}$; for finite dimensional $F$ this 
space can be identified with $F\otimes 
\Phi(\Pi\tilde {{\cal G}})$.) Equivariant cohomology $F$ 
corresponding to generalized functions (distributions) 
on $\Pi\tilde{{\cal G}}$ is denoted by $H_G^{-\infty}(F)$ 
(see[8]). If ${\cal G}$ (and therefore $\tilde {{\cal G}}$) 
is an infinite-dimensional Lie algebra with semi-infinite 
structure, one can define also semi-infinite equivariant 
cohomology of $F$ replacing $\Phi(\Pi\tilde {{\cal G}})$ 
with a Fock space ${\cal F}$ constructed by means of 
$\tilde {{\cal G}}$ (see [12]). In the definition of 
semi-infinite equivariant cohomology one should consider 
$\tilde {{\cal G}}$-semivariants of $F\otimes {\cal F}$ in 
the sense of [9] (instead of $\tilde {{\cal G}}$-invariants 
used in the standard definition of equivariant cohomology.)

   The definition of $G$-equivariant cohomology can be applied 
to every $G^{\prime}$-module. In other words, it can be applied 
to every differential $\tilde {G}$-module, i.e. to 
$\tilde {G}$-module $F$ equipped with an odd differential 
that is compatible with the structure of $\tilde{G}$-module. 
(More precisely, if $F$ is considered as a linear $Q$-manifold 
the map $\tilde {G}\times F\rightarrow F$ that determines an 
action of $\tilde {G}$ on $F$ should be compatible with 
$Q$-structures on $\tilde {G}\times F$ and $F$.) In particular, 
if a group $G$ acts on a manifold $M$ the group $\tilde {G}$ 
acts on the manifold $\tilde M$ and therefore the space 
$\Omega (M)$ of differential forms of $M$ (= the space of 
functions on $\tilde {M}$) can be considered as a differential 
$\tilde {G}$-module. Equivariant cohomology of this module is 
called equivariant cohomology of the $G$-manifold $M$. (If $M$ 
is a supermanifold there are various versions of this definition 
because we can consider different spaces of functions of 
$\tilde M$. Similar remark can be made in the case when $G$ is 
a supergroup.)

  Let us come back to the definition of "string amplitudes". 
We consider $Q$-algebra $(\tilde {P}_n,F,\beta _n)$, where 
$\tilde {P}_n$ constitute a $Q$-EMO with respect to the group 
$\tilde {G}$, $F$ is a differential $\tilde {G}$-module, 
$\beta _n$ are compatible with action of  $\tilde {G}$ and 
with $Q$. (It is not necessary to assume that this $Q$-algebra 
is obtained as a $Q$-extension of an algebra over $P_n$.) In 
other words one can say that $\tilde {P}_n$ constitute an EMO
with respect to the group $G^{\prime}$, a superspace $F$ is 
a $G^{\prime}$-module and the maps 
$\beta _n:\tilde {P}_n\rightarrow F^{\otimes n}$ are compatible 
with $G^{\prime}$-action. If $A_1,...,A_n\in {\cal A}$ are 
physical states (elements of equivariant cohomology of $F$) 
we consider an expression

\begin {equation}
<\hat {A}_1\otimes ...\otimes \hat {A}_n,\beta _n(x)>
\end {equation}
where $x\in \tilde {P}_n$ and $\hat {A}_k$ stands for a 
representative of cohomology class $A_k\in {\cal A}$. It 
is easy to check that (12) determines an element of 
$G$-equivariant cohomology of $P_n$ and that this element 
does not depend on the choice of representatives 
$\hat {A}_k$ in the classes $A_k$. To get "string 
amplitudes" we should have a linear functional on 
$G$-equivariant cohomology of $P_n$ (a kind of integration). 
It is important to emphasize that this construction can be 
applied also in the case when physical states are defined 
by means of semi-infinite equivariant cohomology.

  One can hope to obtain non-perturbative formulation of string 
theory applying the above consideration to the case when 
$P_n=Gr(H^n)$ for appropriate choice of (super)space $H$. 
This hope is based, in particular, on the relation between 
equivariant cohomology of Grassmannian and cohomology of 
moduli spaces of conformal manifolds. To explain this relation 
we should remind some facts about equivariant cohomology. Let 
us assume that a connected compact abelian group $T$ (a torus) 
acts on $M$. Then we can consider equivariant cohomology 
$H_T(M)$ as a module over the polynomial ring 
${\bf C}[\Omega _1,...,\Omega _r]$ where $r=$dim$T$. The 
ring ${\bf C}[\Omega _1,...,\Omega _r]$ can be considered 
as a ring $\Phi$(Lie $T$) of polynomial functions on the 
Lie algebra of $T$.
  
   If $T$ acts on $M$ transitively then $H_T(M)$ can be 
identified with  cohomology of $M/T$ with coefficients in 
the ring $\Phi$(Lie $S$) of polynomial functions on the 
subalgebra of Lie $T$ consisting of elements $t\in T$ 
obeying $t(x)=0$ for fixed point $x\in M$. (In other 
words Lie $S$ is the stabilizer of $x\in M$.) We see 
that the "size" of $H_T(M)$ is determined
by the size of the stabilizer. If the action of $T$ on 
$M$ is not transitive it follows from so called 
localization theorems that the "contribution" of the 
point $x\in M$ to the equivariant cohomology is governed 
by the stabilizer of $x$. In particular, the "free part" 
(the rank) of ${\bf C}[\Omega _1,...,\Omega _r]$-module 
$H_T(M)$ is determined by the fixed points of the action of $T$. 
   
   We can try to apply formally the above statements to the 
action of infinite-dimensional abelian group $\Gamma ^m$ on 
$Gr(H^m)$ where $H=L^2(S^1)$. We will see that 
$\Gamma ^m$-equivariant cohomology of $Gr(H^m)$ can be 
expressed in terms of the Krichever locus and that this 
cohomology is closely related to the cohomology of moduli 
spaces of conformal manifolds. (Recall that Krichever 
locus consists of points having large stabilizers in 
$\Gamma ^m$ and that the space of orbits of $\Gamma ^m$ 
in the Krichever locus $P_m(\Gamma)$ can be identified 
with the moduli space $P_m$ used in the definition of CFT.)
   
   Of course, it is not clear that the statements proved for 
compact groups can be applied to a non-compact group $\Gamma$. 
However, I was able to prove that many results of the theory of 
compact transformation groups can be transferred to non-compact 
case if we understand equivariant cohomology as semi-infinite 
equivariant cohomology [12].
      
{\bf  Appendix.

Isotropic Grassmannian.}
\vskip .1in
  Let us consider a Hilbert space ${\cal H}$ provided with an
antiunitary involution $f\rightarrow f^*$. We will equipp the 
direct sum ${\cal H}^2$ of two copies of ${\cal H}$ with 
semi-infinite structure considering the first copy as $H_+$, 
the second copy as $H_-$ and defining an operator $K$ by the formula

$$K(f,g)=(-g,f)$$

Notice that $K$ is not an involution. However, as we mentioned, 
we can use $K$ to define the spaces $Gr({\cal H}^{2m})$ and the 
maps $\nu _{m,n}: Gr({\cal H}^{2m})\times Gr({\cal H}^{2n}) 
\rightarrow Gr ({\cal H}^{2(m+n)})$ and $\sigma ^{(m)}: Gr({\cal H}^{2m})\rightarrow Gr ({\cal H}^{2(m-2)})$. We will exclude the 
condition 3) from the definition of MO; then these data constitute a MO.

  A linear subspace $V\subset {\cal H}^2$ is called isotropic 
if for every two points $(f,g)\in V,\  (f^{\prime},g^{\prime}) 
\in V$ we have $(f,g^{\prime})+(g,f^{\prime})=0$ (here ( , ) 
denotes bilinear inner product: $(\varphi ,\psi)=<\varphi ,\psi ^*>)$.

   Isotropic Grassmannian $IGr({\cal H})$ can be defined as 
a subset of $Gr({\cal H}^2)$ consisting of isotropic subspaces. 
Giving an obvious definition of isotropic subspace of 
${\cal H}^{2m}$ one can define also $IGr({\cal H}^m)$ as a 
subset of $Gr({\cal H}^{2m})$. It is easy to check that the 
direct sum of isotropic subspaces is again an isotropic
subspace and that the map $\sigma ^{(m)}: Gr({\cal H}^{2m}) 
\rightarrow Gr({\cal H}^{2(m-2)})$ transforms an isotropic
subspace into isotropic subspace. In other words, the 
spaces $IGr({\cal H}^m)$ constitute a MO (a sub MO of the 
MO $Gr({\cal H}^{2m})$). The decomposition $Gr({\cal H}^{2m}) 
=\cup Gr^{(k)}({\cal H}^{2m})$ induces decomposition 
$IGr({\cal H}^m)=\cup IGr^{(k)}({\cal H}^m)$, hence we 
consider the MO $IGr({\cal H}^m)$ as graded MO. In 
particular, we can say that the spaces $IGr^{(0)}({\cal H}^m)$ 
also constitute a MO.

  It is important to notice that, conversely, usual 
Grassmannian can be embedded into isotropic Grassmannian. 
Let us assume that ${\cal H}$ is equipped with a 
semi-infinite structure $({\cal H}={\cal H}_+\oplus
{\cal H}_-)$ and that antiunitary involution 
$f\rightarrow f^*$ transforms ${\cal H}_+$ into 
${\cal H}_+$ and ${\cal H}_-$ into ${\cal H}_-$. 
Then for every linear subspace $V\subset {\cal H}$ 
we can construct an isotropic subspace 
$\rho (V)\subset {\cal H}^2$ as a direct sum of 
subspaces $\pi _1V$ and $\pi _2V^{\perp}$, where 
$\pi _1:{\cal H}\rightarrow {\cal H}^2$ transforms 
$f\in {\cal H}$ into $(\pi _+f,\pi _-f)$ and $\pi _2: 
{\cal H}\rightarrow {\cal H}^2$ transforms $f\in {\cal H}$ 
into $(\pi  _-f,\pi _+f)$. (Here $\pi_{\pm}:{\cal H}\rightarrow 
{\cal H}_{\pm}$ are orthogonal projections and the orthogonal 
complement $V^{\perp}$ is taken with respect to bilinear inner 
product $(\varphi ,\psi)=<\varphi ,\psi ^*>$.) It is easy to 
check that for $V\in Gr({\cal H})$ we have 
$\rho (V) \in IGr({\cal H})$; hence we embedded $Gr({\cal H})$ 
into $IGr({\cal H})$. Moreover, one can verify that index 
$\rho (V)=0$ therefore we can say that we embedded 
$Gr({\cal H})$ into $IGr^{(0)}({\cal H})$.

 Let us consider a Hilbert space ${\cal H}$ with 
antiunitary involution $f\rightarrow f^*$. We define 
the space ${\cal F}$ as the space of antiholomorphic 
functionals $\Phi (a^*)$ on $\Pi {\cal H}$ obeying the
condition

\begin {equation}
\int \Phi (a^*)\Phi^*(a)e^{-(a,a^*)}dada^*<\infty .
\end {equation}
Here $\Pi{\cal H}$ stands for the superspace obtained from 
${\cal H}$ by means of parity reversion; $\Phi ^*(a)= 
(\Phi(a^*))^*$. In other words, ${\cal F}$ consists of 
elements of infinite-dimensional  Grassmann algebra generated 
by ${\cal H}$, i. e. of formal expressions

$$\Phi (a^*)=\sum_n{1\over (n!)^{1/2}}\int \Phi _n(x_1,...,x_n)a^*(x_1)...a^*(x
_n)d^nx,$$

where $\Phi _n(x_1,...,x_n)$ are antisymmetric and

$$\sum \int |\Phi _n|^2d^nx<\infty .$$

(We consider ${\cal H}$ as the space of square integrable 
functions depending on $x\in S$, where $S$ is a measure 
space; involution is realized as complex conjugation. 
This restriction is not essential, but it permits us to 
simplify notations.) It is well known that ${\cal F}$
can be considered as fermionic Fock space. (See[10]; we 
follow the notations of this book. A rigorous explanation 
of the meaning of infinite-dimensional integral in (13) 
also can be found in [10].)

   Operator of multiplication on $\int f(x)a^*(x)dx$ will 
be denoted by $a^+(f)$ and operator 
$\int f(x){\delta \over \delta a^*(x)}dx$ by $a(f)$. These 
operators obey canonical anticommutation relations (Clifford 
algebra relations) $[a(f),a(f^{\prime})]_+=[a^+(f),a^+ 
(f^{\prime})]_+=0,\ [a(f),a^+(f^{\prime})]_+=(f,f^{\prime})$. 
The functional $\Phi_0=1$ can be considered as vacuum vector; 
it obeys $a(f)\Phi _0=0$ for all $f\in {\cal H}$. Operator 
$a^+(f^*)$ is adjoint to $a(f)$ with respect to Hermitian inner 
product 

$$<\Phi _1,\Phi _2>=\int \Phi _1(a^*)\Phi _2^*(a)e^{-(a,a^*)}dada^*.$$

For every vector $\Psi \in {\cal F}$ we define a linear 
subspace Ann $\Psi \subset {\cal H}^2$ consisting of such 
pairs $(f,g)\in {\cal H}^2$ that

$$(a(f)+a^+(g))\Psi =0$$

  It is easy to check that the subspace Ann $\Psi$ is isotropic. 
One can prove that for every $V\in IGr^{(0)}({\cal H})$ there 
exists a vector $\Psi =\Psi_V\in {\cal F}$ obeying $V=$Ann $\Psi$; 
this vector is unique up to a factor. (We assume that Grassmannian 
is defined by means of Hilbert-Schmidt operators; this is 
essential for the validity of the above statement.) To give a proof 
we represent $V\in IGr^{(0)}({\cal H})$ as an image of a linear 
map ${\cal H}\rightarrow {\cal H}^2$ transforming 
$\varphi \in {\cal H}$ into $(A\varphi ,B\varphi)$. Here
$A$ is a Fredholm operator of index $0$, $B$ is a  Hilbert-Schmidt
operator. The condition $V\subset $Ann$\Psi$ means that for every
$\varphi \in {\cal H}$ we have

\begin {equation}
(\sum A_{kl}\varphi _l{\partial \over \partial a_k^*}+\sum
B_{kl}\varphi _la_k^*)\Psi =0
\end {equation}
(We have chosen a basis in ${\cal H}$.) The condition that the 
space $V$ is isotropic means that

\begin {equation}
\sum _k(A_{kl}B_{kr}+B_{kl}A_{kr})=0.
\end {equation}
   Without loss of generality we can assume that the matrix 
$A_{kl}$ is diagonal; moreover, non-zero entries can be  
taken equal to $1$. We assume that $A_{ii}=0$ for 
$i\leq s,\ A_{ii}=1$ for $i>s$. Then the equations (14) 
and the condition (15) take the form

\begin {equation}
(\sum _kB_{kl}a_k^*)\Psi=0\ \ \ \ {\rm for } \  l\leq s,
\end {equation}

\begin {equation}
({\partial \Psi \over \partial a_l^*}+\sum _kB_{kl}a_k^*)\Psi\ 
\ \ \ \ {\rm for }\  l>s.
\end {equation}

$$B_{lr}+B_{rl}=0 \ \ \ {\rm if }\   l>s,\  r>s$$

$$B_{lr}=0 \ \ \  {\rm if }\   l>s,\  r\leq s \ \ \  {\rm or }\   
l\leq s,\  r>s.$$

  The solution to equations (16),(17) is unique (up to a factor) 
and can be written in the form
\begin {equation}
\Psi=C\cdot \Pi_{l\leq s} \delta (\sum _kB_{kl}a_k^*)e^{-{1\over 2}\sum
_{k,l\geq s}a_l^*B_{kl}a_k^*}
\end {equation}
or in the form

\begin {equation}
\Psi=C\cdot \Pi_{l\leq s} (\sum _kB_{kl}a_k^*)e^{-{1\over 2}\sum
_{k,l\geq s}a_l^*B_{kl}a_k^*}.
\end {equation}
These two forms are equivalent because $\delta (a_k^*)=a_k^*$. 
(In other words, we have $\int \delta (a_k^*)\varphi 
(a^*,a)dada^*=\int a_k^*\varphi  (a^*,a)dada^*$ for 
every $\varphi$.) Notice, that we proved a little bit 
more than claimed. Namely, it follows from the proof 
that for $V\in IGr^{(0)}({\cal H})$ there exists unique 
(up to a factor) vector $\Psi$ obeying Ann $\Psi \supset V$ 
and that we have Ann $\Psi =V$ for this vector.

 In the case when the operator $A$  is invertible we can 
write the functional $\Psi =\Psi _V$ in the form

$$\Psi _V=Ce^{-{1\over 2}(a^*,A^{-1}Ba^*)}.$$

(It is clear that $\Psi _V$ satisfies (14).) If the operator 
$A$ has even number of zero modes we can represent 
$V\in IGr^{(0)}({\cal H})$ as a limit of  $V_n\in IGr^{(0)}
({\cal H})$ in such a way that $V_n$ is an image of a map 
${\cal H}\rightarrow {\cal H}^2$ transforming
$\varphi \in {\cal H}$ into $(A_n\varphi ,B_n\varphi )$, 
where $A_n$ is an invertible operator, $A_n-1$ belongs to 
the trace class and $B_n$ is a Hilbert-Schmidt operator. 
Using this representation one can write
$\Psi _V$ as a limit of functionals

$$(\det A_n)^{1/2}e^{-{1\over 2}(a^*,A_n^{-1}B_na^*)}.$$

  We constructed a map $IGr^{(0)}({\cal H})\rightarrow {\cal F}$
(defined up to a factor).  It is easy to check that applying this
construction $IGr^{(0)}({\cal H}^m)$ we obtain an algebra over MO
$IGr^{(0)}({\cal H}^m)$.

  The definition of  isotropic Grassmannian $IGr({\cal H})$ can be
easily generalized to the case when ${\cal H}={\cal H}_0\oplus {\cal
H}_1$ is a complex Hilbert $Z_2$-graded space with antiunitary
involution $f\rightarrow f^*$. (Bilinear inner product $(f,g)=<f,g^*>$
should be symmetric in the sense of superalgebra: $(g,f)=(f,g)$ if $f$
and $g$ are even, $(g,f)=-(f,g)$ if $f$ and $g$ are odd.) Almost all
consideration above can be repeated with some changes. In
particular, one can relate $IGr({\cal H})$ to the Fock space ${\cal F}$
defined as a space of antiholomorphic functionals $\Phi (a^*)$  on $\Pi
{\cal H}$ satisfying the condition (13). Let us assume that the space
${\cal H}$ is realized as a space of functions on measure space $S$
taking values in ${\bf C}^{p|q}$. (In other words elements of ${\cal
H}$ are functions $f(x,\alpha )$ of $x\in S$ and discrete index
$\alpha$; we assume that $f(x,\alpha )$ is even for $1\leq \alpha \leq
p$ and $f(x,\alpha )$ is odd for $p+1\leq \alpha \leq p+q$. We will say
that $\alpha $ is a superindex, taking $p$ even and $q$ odd values.)
One can represent an element of ${\cal F}$ as an expression on the form

$$\Phi (a^*)=\sum{1\over (n!)^{1/2}}\sum _{\alpha _1,...\alpha _n}\Phi
_n(x_1,\alpha _1,...,x_n,\alpha _n)a^*(x_1,\alpha _1)...a^*(x_n,\alpha
_n)d^nx$$
obeying

$$\sum _{\alpha _1,...\alpha _n}|\Phi _n(x_1,\alpha _1,...,x_n,\alpha
_n)|^2d^nx<\infty.$$

(Here $x_i\in S,\  \alpha _i$ is a superindex taking $p$ even values
and $q$ odd values, the function $\Phi$ is antisymmetric with respect
to transposition $(x_i,\alpha _i)$ and $(x_j,\alpha _j)$ if $\alpha _i$
and $\alpha _j$ are even indices and symmetric in all other cases.)
Multiplication by $\sum _{\alpha}\int f(x,\alpha )a^*(x,\alpha )dx$
determines an operator $a^+(f)$ acting on ${\cal F}$ and linearly
depending of $f\in {\cal H}$. Operators $a^+(f)$ together with
operators $a(f)=\sum _{\alpha }\int f(x,\alpha ){\delta \over \delta
a^*(x,\alpha )}dx$ generate a superanalog of Clifford algebra. We can
repeat the definition of linear subspace Ann $\Psi \subset {\cal H}^2$
and prove that Ann $\Psi$ is isotropic. (Here $\Psi \in {\cal F}$.)
Representing $V\in IGr^{(0)}({\cal H})$ as an image of linear map
$\varphi \rightarrow (A\varphi ,B\varphi )$, where $A$ is Fredholm
operator of index $0$, $B$ is a Hilbert-Schmidt operator we obtain a
condition of isotropicity and a condition that $V\subset$Ann$\Psi$.
These conditions coincide with (15) and (14) correspondingly (up to
irrelevant signs). For generic $V$ we can assume that $A_{kl}$ is
diagonal, $A_{ii}=0$ for $i\leq s,\  A_{ii}=1$ for $i>s$. Then we can
solve the analog of Equation (16) and obtain the expression (18) for
the functional $\Psi =\Psi _V$; the solution is unique up to a factor.
(Of course, (18) is not equivalent to (19) in general case.) We see
that  the functional $\Psi$ can be considered as an element of ${\cal
F}$ only in the case when the odd-odd block of the matrix $A$ is
invertible. A map $V\rightarrow \Psi _V$ can be considered as a
generalized function on $IGr^{(0)}({\cal H})$ with values in ${\cal F}$.
(More precisely, this function takes values in appropriate extension
of the Fock space.)
Analogously we obtain a generalized function $IGr^{(0)}({\cal
H}^m)\rightarrow {\cal F}^{\otimes m}$. It is easy to prove that this 
function can be considered as a (generalized) algebra over 
$IGr^{(0)}({\cal H}^m)$.

  If $V\in IGr^{(0)}({\cal H})$ is represented as an image of 
a map $\varphi \rightarrow (A\varphi, B\varphi )$, where $A$ 
is invertible operator, $A-1$ belongs to trace class and $B$ 
is a Hilbert-Schmidt operator, we can write
\begin {equation}
\Psi _V=(\det A)^{1/2}e^{-{1\over 2}(a^*,A^{-1}Ba^*)}
\end {equation}
Of  course, $\det$ here and further stands for superdeterminant 
(Berezinian). For general $A$ one can get $\Psi_V$ taking a limit 
in (20), as we explained above in the case when ${\cal H}$ is a 
ordinary Hilbert space. Of course all our consideration determine 
$\Psi _V$ only up to a factor.
  Using the  embedding of ordinary Grassmannian into isotropic 
Grassmannian we can construct a function $V\rightarrow \Psi _V$ 
defined for $V\in Gr(H)$ and taking values in Fock space (if $H$ 
is a superspace then $\Psi _V$ takes values in an appropriate extension
of Fock space). This function  determines a (generalized) algebra
over Grassmannian MO. One 
can define $\Psi _V$ only up to a factor; however one can give 
an unambiguos definition of $\Psi _V$ on $\tilde {Gr}(H)$ 
naturally embedded into $Gr(\tilde {H})$ (see below). We 
mentioned already that the maps $\Psi _V$ determine an 
algebra over MO $Gr(H^k)$. One can reformulate the statement 
above saying that the maps $\Psi _V$ on $\tilde {Gr}(H^k)\subset 
Gr(\tilde {H}^k)$ specify a $Q$-extension of this algebra 
(a $Q$-algebra over $Q$-MO $\tilde { Gr}(H^k)$).

One should emphasize, that our consideration of 
super Grassmannian and its connection with Fock space was
neither rigorous, nor complete. More detailed treatment
of related questions can be found in [17], [18], [19].

  Let us discuss briefly more general way to construct 
generalized algebras over Grassmannian MO's. Every algebra 
over $Gr(H^n),\  H=L^2(S^1)$ determines an algebra over MO 
$P_n$ of surfaces with disks (a conformal field theory), 
because $P_n$ is embedded into $Gr(H^n)$ by means of 
Krichever construction. Conversely, a conformal field 
theory can be extended to an algebra over $Gr(H^n)$ if 
this theory is defined by means of quadratic Lagrangian 
(free fermions or bosons, $bc$-sistem, $\beta \gamma$-system 
or any combination of these theories). If a CFT 
$(P_n,E,\alpha _n)$ has non-vanishing central charge then 
$\alpha _n$ should be considered not as a map 
$P_n\rightarrow E$, but as a section of a bundle 
$E\otimes (\det)^k$ where $\det$ stands for so called 
determinant bundle over $P_n$. It follows from well 
known results about determinant bundles over $P_n$ 
and $Gr(H^n)$ that an extension of a conformal field 
theory having vanishing central charge to the 
Grassmannian is an algebra in strict sense 
(i.e. there exists an unambiguos definition of a 
vector corresponding to an element $V\in  Gr(H^m)$). 
  
  Using the same idea one can check the above mentioned 
fact that $\Psi _V$ can be defined on $\tilde Gr(H)$ 
unambiguously.
  
{\bf Acknowledgments.} I am indebted to L. Dickey, M. Duflo,
 M. Kontsevich, 
M. Mulase, M. Vergne, A. Voronov and B. Zwiebach for useful 
discussions.
  
{\bf References.}

1. Segal, G., Wilson, G.: Loup Groups and Equations of KdV type. 
IHES Publ Math., 61,5-65 (1985)

2. Mulase, M., Cohomological structure in soliton equations and 
Jacobian varieties, J. of Diff. Geom., 19, 403-430 (1984)

3. Mulase, M., Schwarz, A.: (in preparation)

4. Getzler, E., Kapranov, M.: Modular operads, preprint

5. Getzler, E.: Two-dimensional topological gravity and equivariant 
cohomology. Comm. Math. Phys., 163, 473-489 (1994)

6. Schwarz, A.: Geometry of Batalin-Vilkovisky quantization. 
Comm. Math. Phys., 155, 249-260 (1993), Semiclassical 
approximation in Batalin-Vilkovisky formalism. Comm. 
Math. Phys., 158, 373-396 (1993)

7. Sen, A., Zwiebach, B.: Quantum background independence of 
closed string field theory. Nucl. Phys., B423, 580 (1994)

8. Duflo, M., Kumar, S., Vergne, M.: Sur la cohomologie 
equivariante des varietes differentiables. Asterisque, 215 (1993)

9. Voronov, A., Semi-infinite homological algebra. Inv. 
Math., 113,103-146 (1993)

10. Berezin, F.: The method of second quantization. Academic 
Press, New York-London (1966)

11. Hsiang, W.Y., Cohomology theory of topological 
transformation groups. Springer-Verlag, Berlin-Heidelberg-New 
York,1975

12. Schwarz, A., (in preparation)

13. Atiyah, M., Bott, R., The moment map and equivariant cohomology.
Topology,23, 1-28 (1993)

14. Berline, N., Geltzler, E., Verline, M., Heat kernels and Dirac 
operators. Springer, Berlin,1991

15. Witten, E., Two-dimensional gauge theories revisited. J. of 
Geom. Phys., 9, 303 (1992)

16. Schwarz, A., Zaboronsky, O., Supersymmetry and localization. 
hep-th 95 11112 (to be published in CMP)

17. Schwarz, A. Fermionic string and universal moduli space
Nucl. Phys. B317, 323 (1989)

18.Alvarez-Gaume, L., Nelson, Ph., Gomez, C., Sierra, G., Vafa, C.,
Fermionic strings in the operator formalism, Nucl. Phys. B311, 333
(1989)

19. Dolgikh, S., Schwarz, A. , Supergrassmannians, super tau-functions
and strings, in: Physics and Mathematics of Strings, pp 231-244.
Editors L. Brink, D. Friedan, A. Polyakov, World Scientific,
  Singapore, 1990

\end {document}